\documentclass[twocolumn,showpacs,preprintnumbers,amsmath,amssymb]{revtex4} 
\input psfig.sty 
\def \be {\begin{equation}} 

\def \ee {\end{equation}} 
\def \bea {\begin{eqnarray}} 
\def \eea {\end{eqnarray}}

\usepackage{graphicx,subfigure}
\usepackage{dcolumn}
\usepackage{bm}
\usepackage{epsfig}

\begin{document} 

\title{Supernovae as probes of cosmic parameters: estimating the bias from under-dense lines of sight}

\author{V. C. Busti$^1$} \email{vinicius.busti@uct.ac.za} 

\author{R. F. L. Holanda$^{2,3}$} \email{holanda@uepb.edu.br}

\author{C. Clarkson$^1$} \email{chris.clarkson@uct.ac.za} 

\vskip 0.5cm \affiliation{$^1$Astrophysics, Cosmology \& Gravity Center (ACGC), and Department of Mathematics and Applied Mathematics, 
University of Cape Town, Rondebosch 7701, Cape Town, South Africa \\ $^{2}$ Departamento de F\'isica, Universidade Estadual da Para\'iba,
58429-500, Campina Grande -- PB, Brasil \\ $^3$ Departamento de F\'isica, Universidade Federal de Campina Grande, 58429-500, 
Campina Grande -- PB, Brasil } 


\begin{abstract} 
\noindent 

Correctly interpreting observations of sources such as type Ia supernovae (SNe Ia) require knowledge of the power spectrum of matter on AU scales~-- which is very hard to model accurately. Because under-dense regions account for much of the volume of the universe, light from a typical source probes a mean density significantly below the cosmic mean. The relative sparsity of sources implies that there could be a significant bias when inferring distances of SNe Ia, and consequently a bias in cosmological parameter estimation. While the weak lensing approximation should in principle give the correct prediction for this, linear perturbation theory predicts an effectively infinite variance in the convergence for ultra-narrow beams.  We attempt to quantify the effect typically under-dense lines of sight might have in parameter estimation by considering three alternative methods for estimating distances, in addition to the usual weak lensing approximation. We find in each case this not only increases the errors in the inferred density parameters, but also introduces a bias in the posterior value.

\end{abstract} 

\maketitle 

\section{Introduction}

Over the last decade exquisite observations of several 
cosmological probes (e.g. type Ia supernovae (SNe Ia) \cite{sn1998},
cosmic microwave background (CMB) temperature anisotropies \cite{cmb_ani}, baryon acoustic oscillations (BAOs) \cite{bao_meas}) has led the emergence of the so-called standard model in cosmology, known as the $\Lambda$CDM model, and the determination of cosmological parameters with a precision of a few percent. However, the nature of the basic components of the model, the cold dark matter and the cosmological constant, is still unknown and there are a large variety of alternatives~\cite{gen_reviews}.

Of these observables, light from SNe Ia probe the fully non-linear regime because they traverse structure over large distances with a beam which is very narrow. The distance to the SNIa is affected by the usual weak lensing of large-scale structure, an effect which is now being detected~\cite{Smith:2013bha}. What is the full effect of inhomogeneities along the line of sight to an SNIa? 

Ray-tracing techniques were applied to cosmological N-body simulations
to analyze the magnitude of the effects of the inhomogeneities and found small deviations \cite{nbody}, implying that
the inhomogeneities are compensated along the line of sight. If this assumption is relaxed, very different results are achieved. If light travels preferentially through an underdense medium, as it seems to be the case in the web structure we see, different distances are derived. This point was raised by Bolejko \cite{bol_dr_2011} and Meures \& Bruni \cite{meures_2012}. Moreover, in the case that some lines of sight are blocked due to opaque structures of high density,
underdense lines would be preferred \cite{futamase_2009,mattsson}.

The size of the light beam is the key issue. For beams with sizes on arcminute scales, the dispersion in the Hubble diagram due to matter fluctuations
can be corrected \cite{corr_lens} due to the shearing of images. As a supernova beam has an angular size of $\sim$ $10^{-7}$ arcsec at a redshift
$z \sim 1$, this correction cannot be applied, since shear maps are smoothed on arcminute scales \cite{dalal}. Furthermore, the typical size 
of a beam in an N-body simulation is around hundreds of kiloparsecs, while a supernova has 1 a.u. size ($\sim10^{-9}$\,kpc). Non-linear terms in the 
mean magnification must be considered, which can shift the mean by a non-negligible amount. This is extremely dependent on the matter distribution 
along the line of sight on scales not covered by perturbation theory or in N-body simulations. In particular, perturbation theory predicts a variance which diverges as the beam size becomes small which means it loses predictive power~\cite{clarkson12}. It was shown in~\cite{clarkson12} that narrowing the beam size in N-body simulations shifts the probability distribution function dramatically, implying that most narrow beams probe very underdense  lines of sight which are compensated by relatively few of high density. Of course, these simulations can only probe beam sizes which are many of orders of magnitude larger than required, so one can only speculate as to the actual probability distribution function. 

Given this uncertainty in knowing the correct distribution function to use, there is uncertainty in modelling the magnification of such narrow beams, which relies on the matter density and expansion rate along the beam.  
In this paper we follow an observational approach. We choose four different approximations to probe  inhomogeneities
with observations \cite{other}. 
Three of them can change the cosmological parameters by several percent:
the Dyer--Roeder (DR) approximation \cite{dr}, the weak lensing approximation with uncompensated density along the line of sight \cite{bol_dr_2011}, 
and the flux-averaging approximation \cite{flux_av}. We also propose a new approximation which takes into account the different expansion rates along the line of sight. 

In our analyses we use two samples of SNe Ia.  The first is the Union2.1 compilation data \cite{union2.1}, comprising 580 SNe Ia calibrated with the SALT2 light curve fitter \cite{salt2}.
In this sample we add a high-redshift supernova SCP-0401 detected at $z=1.713$ by Rubin {\it et al.} \cite{rubin}. For convenience,
we call the set of 581 SNe as the Union2.1 compilation data.
The second sample comprises 288 SNe Ia of the First-Year Sloan Digital Sky Survey II \cite{sdss}, where we consider the data calibrated
with the MLCS2k2 light curve fitter \cite{mlcs}. We call this sample  the SDSS compilation data. 

A possible way to disentangle the effects
of the inhomogeneities is to consider joint analyses with a sample with a different degeneracy in the parameter space. 
So, we also consider 19 $H(z)$ measurements from differential age of 
passively evolving galaxies \cite{hzmeasurements}. In order to deal with a tension between different measurements of the Hubble constant $H_0$,
we also consider two measurements for $H_0$: $73.8\pm 2.4$ km s$^{-1}$ Mpc$^{-1}$ \cite{H0riess} and $68.0 \pm 2.8$ km s$^{-1}$ Mpc$^{-1}$ 
\cite{H0CR}, where the latter is in agreement with the latest measurement from {\it PLANCK} \cite{planck}.

The paper is organized as follows. In Sec. \ref{sec_lp} we give a brief overview of light propagation in a general spacetime.
In Sec. \ref{sec_th} the different approximations to deal with the light propagation are presented.
In Sec. \ref{sec_res} we present the samples used and the results obtained from the statistical analyses. We finish the paper
in Sec. \ref{sec_conc} with the conclusions. 

\section{Light Propagation}
\label{sec_lp}

In this section we give a brief overview of light propagation in a generic spacetime. The idea is to derive the Sachs optical equations \cite{sachs}
used in the derivation of the approximations discussed in the next section. We refer the reader to \cite{clarkson12} for detailed
explanations.

To describe light propagation the geometric optics approximation \cite{SEF92} is assumed. Light rays are bundles of irrotational null 
geodesics $x^\mu(v,s)$, where $v$ is the affine parameter and $s$ labels the geodesics, so the tangent vector $ k^\mu = {d x^\mu}/{d v}$ obeys 

\begin{equation}
k^\mu k_\mu = 0, \,\, k^\nu \nabla_\nu k_\mu = 0, \,\, \nabla_{[\mu}k_{\nu]}=0.
\label{nullgeod}
\end{equation}

The connecting vector $\eta^\mu = d x^\mu/d s$ which links neighbouring geodesics and gives the physical shape of the bundle 
satisfies the geodesic deviation equation

\begin{equation}\label{gde}
 k^\alpha k^\beta \nabla_\alpha \nabla_\beta \eta^\mu = {R^\mu}_{\nu\alpha\beta} k^\nu k^\alpha\eta^\beta.
\end{equation}

We can project the above equation in a screen space orthogonal to the ray direction as

\begin{equation}\label{eq_eta}
 \frac{d^2}{d v^2}\eta_a = {\cal R}_{ab}\eta^b,
\end{equation}
where ${\cal R}_{ab}={R}_{\mu\nu\alpha\beta}k^\nu k^\alpha n_a^\mu n_b^\beta$ is the screen projection of the Riemann tensor, and $n^\mu_a$ ($a=1,2$) are 
unit vectors spanning the screen space. ${\cal R}_{ab}$ can be decomposed in the following way

\begin{equation}
 {\cal R}_{ab} =
  \left(\begin{array}{cc}\Phi_{00} & 0 \\ 0 & \Phi_{00}\end{array} \right)
  +
  \left(\begin{array}{cc} - {\rm Re}\,\Psi_0 & {\rm Im}\,\Psi_0 \\
  {\rm Im}\,\Psi_0 & {\rm Re}\,\Psi_0\end{array} \right)
\end{equation}
with
\begin{equation}\label{ricweyl}
 \Phi_{00}=-\frac12 R_{\mu\nu}k^\mu k^\nu,\quad
 \Psi_0=-\frac{1}{2}C_{\mu\nu\alpha\beta}m^\mu
 k^\nu m^\alpha k^\beta,
\end{equation}
and $m^\mu \equiv n_1^{\mu} - {\rm i}  n_2^{\mu}$. $\Phi_{00}$ is called the Ricci focusing and $\Psi_0$ is called the Weyl focusing. The former
is generated by matter inside the beam while the latter is generated by matter outside the beam that induces a non-vanising Weyl tensor inside
the beam.

We can manipulate Eq. (\ref{eq_eta}) and express it as a function of the optical scalars $\hat{\theta}$ and 
$\hat{\sigma}=\hat{\sigma}_1+i\hat{\sigma}_2$,
called the null expansion and the null shear respectively, which gives the Sachs equations \cite{sachs} 

\begin{eqnarray}
 \frac{d\hat\theta}{d v} + \hat\theta^2 + \vert\hat\sigma\vert^2 &=& \Phi_{00}, \label{e.1}\\
 \frac{d\hat\sigma}{d v} +2 \hat\theta \hat\sigma &=& \Psi_0 ,\label{s2} \\
 \hat\theta \equiv {1\over2}\nabla_\mu k^\mu,~~|\hat\sigma^2| & \equiv & {1\over2}\nabla_\mu k_\nu \nabla^\mu k^\nu-\hat\theta^2. \label{s3}
\end{eqnarray}

The null expansion is related to the cross-sectional area $A$ of the light rays as \cite{poisson}

\begin{equation}
\hat\theta = \frac{1}{\sqrt{A}}\frac{d}{d v} \sqrt{A}.
\end{equation}
Since the angular diameter distance $D_A$ is proportional to $\sqrt{A}$, we obtain

\begin{eqnarray}
\frac{d^2 D_A}{d v^2} &=& - \left(\vert\hat\sigma\vert^2   - \Phi_{00}\right)D_A. 
\end{eqnarray}

A last step is needed before we move to particular approximations, which is the transformation of the affine parameter $v$ to the observable 
redshift $z$. For observers with four-velocity $u^{\mu}$, the redshift is given by

\begin{equation}\label{eq_red}
 1+z(v) =\frac{(k_\mu u^\mu)_{v}}{(k_\mu u^\mu)_{0}}.
\end{equation}
This implies that \cite{cm2010}

\begin{equation}
\frac{d z}{d v} = \frac{d(u^\mu k_\mu)}{dv}  =
(1+z)^2H_\parallel(z,e^{\mu}),
 \end{equation}
where $H_\parallel$ is the observed expansion rate along the line of sight and $e^{\mu}$ is the spatial direction of observation. From these
equations we see that assumptions must be specified about the matter distribution and the expansion rate along the line of sight.
 
\section{Theoretical Models}
\label{sec_th}

We now focus our attention to four approximations that try to take into account the effects of the inhomogeneities. As they have
different premisses it is interesting to see how cosmological parameters are affected by them. Let us describe them starting with the DR approximation.

\subsection{The DR approximation}

The DR approximation \cite{dr} assumes that light can propagate preferentially by underdense lines of sight. This fact, first noted by 
Zel'dovich \cite{zeldovich}, is incorporated in the model
through the introduction of the smoothness parameter $\alpha$ in $\Phi_{00} \rightarrow \alpha\Phi_{00}$. For $\alpha=1$ we have the same amount
of matter as in the homogeneous case, so it is called the filled beam. In the other extreme, if light propagates in vacuum we have $\alpha=0$, 
which is the empty beam. Therefore, for a partial clumping, the smoothness parameter varies between 0 and 1. The approximation also requires
that the shear is zero and the expansion along the line of sight is the same as in the homogeneous case ($H_\parallel=H$).

In short, the DR approximation states that light propagates in a homogeneous universe with less matter inside the beam which is redistributed to clumps.
It is based in the fact that the probability of a typical line of sight to encounter a high-mass halo is low and if this happens strong lensing
can occur, and the supernova would be removed of the statistical analysis, or the supernova is blocked and we would not see it. So, it is a simple
way to quantify a possible bias in the statistical analyses using SNe Ia.

Adopting these assumptions and restricting our attention to a flat $\Lambda$CDM 
model  (see \cite{sereno} for the influence of a quintessence fluid), we derive the DR equation

\begin{eqnarray}
 \frac{d^2 D_A}{d z^2} &+& \left(\frac{d\ln H}{d z}
 +\frac{2}{1+z}\right) \frac{d D_A}{d z}
 \nonumber \\
 &=&-\frac{3}{2}\Omega_{\rm m} 
 \frac{H_0^2}{H^2}(1+z)
 \alpha(z) D_A,
\label{dr_eq}
\end{eqnarray}
where $\Omega_{\rm m}$ is the matter density parameter today and the redshift dependence of $\alpha$ encodes the fact that we expect the Universe
to be more homogeneous in the past \cite{linder,tomita,mortsell}.

In order to compare this approximation with the SNe Ia data, the relation between the luminosity distance and the angular diameter distance 
$D_L=(1+z)^2 D_A$ is used, which is known as the Etherington Principle \cite{etherington}. Writing in terms of the adimensional luminosity
distance $d_L=(H_0/c)D_L$, where $c$ is the speed of light, we have

\begin{equation}\label{angdiamalpha}
 \left( 1+z\right) ^{2}{\cal{F}}
\frac{d^{2}d_L}{dz^{2}} - \left( 1+z\right) {\cal{G}}
\frac{dd_L}{dz} + {\cal{H}} d_L=0,
\end{equation}
which satisfies the initial conditions
\begin{equation}
\left\{
\begin{array}{c}
d_L\left( 0\right) =0, \\
\\
\frac{dd_L}{dz}|_{0}=1.
\end{array}
\right.
\end{equation}
The terms $\cal{F}$, $\cal{G}$ end $\cal{H}$ are functions of the cosmological parameters, expressed in terms of 
the redshift by

\begin{eqnarray}
{\cal{F}}& =& \Omega_{\rm m} + (1-\Omega_{\rm m} )(1+z)^{-3},\nonumber
\\ \nonumber \\ {\cal{G}} &=& \frac{\Omega_{\rm m}}{2}
+2(1-\Omega_{\rm m} )(1+z)^{-3},\nonumber
\\ \nonumber
\\ {\cal{H}} &=& \left[\frac{3\alpha(z)-2}{2}\right]\Omega_{\rm m}
 + 2(1-\Omega_{\rm m})(1+z)^{-3}. \\
\nonumber
\end{eqnarray}

The DR approximation has been criticized by some authors. The first came from photon flux conversation put forward by Weinberg \cite{weinberg76},
where divergence from underdense regions are compensated by convergence of clumpy regions. The question remains open, since following 
works arrived at different conclusions \cite{photon_cons}. More recently, R\"as\"anen \cite{rasanen} questioned the DR approximation,
but relied that the density along the line of sight is compensated and there are no selection effects, which are not necessarily the case
\cite{futamase_2009,mattsson,bol_dr_2011}. In fact, a method was developed in \cite{futamase_2009} to take into account the blocked lines of sight, where halos above
a minimum threshold do not allow light beams to cross them. As a result, one ends up with a specific form for $\alpha(z)$ given by the matter not locked in clumps. As the universe
was more homogeneous in the past, a smaller deviation compared to the standar DR model is derived. Therefore, we will keep our attention to the standard, and most extreme, DR approximation.

\subsection{The weak lensing approximation}

The weak lensing approximation considers perturbations to a homogeneous background where the line element in Newtonian gauge is

\begin{equation}
 d s^2 = a^2(\eta)\left[-(1+2\Phi)d\eta^2 + (1-2\Psi)\gamma_{ij}d x^i d x^j \right],
\end{equation}
$\Phi$ and $\Psi$ are the Bardeen potentials.

The angular diameter distance in this case is \cite{clarkson12,futamase_1989}
\begin{equation}
 D_A= \bar D_A(1+\delta_A),
\end{equation}
where $\bar D_A$ is the homogeneous angular diameter distance and $\delta_A$ is the negative of convergence given as an integral of the homogeneous
distance and the perturbed potentials.

Recently, Bolejko \cite{bol_dr_2011} found out a relation between the weak lensing and DR approximations, firstly discussed by Futamase and Sasaki \cite{futamase_1989}. They provide the same results
when the smoothness parameter has the following form

\begin{equation}
 \alpha(z) = 1 + \frac{\langle\delta\rangle_{1D}}{(1+z)^{5\over4}},
\end{equation}
where $\langle\delta\rangle_{1D}$ is the mean of density fluctuations along the line of sight. 

In this formulation, the standard weak lensing
approach predicts that the inhomogeneities are compensated along the line of sight and $\langle\delta\rangle_{1D}$ is of order $10^{-3}$.
However, that is not necessarily the case, since the density fluctuations are not randomly distributed but
form a cosmic web. In this case the value of $\langle\delta\rangle_{1D}$ can be much higher and very different distances would be derived. From now on we consider light travelling in an underdense medium so that $-1\leq \langle\delta\rangle_{1D} \leq 0$, and we constrain the modulus of it. Note that light travelling in overdense medium 
is possible and was preferred by Union2 compilation data \cite{union2} as shown in \cite{alpham1} for an extended DR approximation, but a correspondence
with weak lensing is not available yet.

\subsection{The flux-averaging approximation}
\label{faa}

The flux-averaging approximation was developed by Wang and collaborators \cite{flux_av} and it is based on the flux conservation of gravitational
lensing as emphasized by Weinberg \cite{weinberg76}. The idea is that considering a large number of standard candles in the same redshift,
the average magnification is one. In this way, the process should reduce the bias produced by gravitational lensing. However, note that
effects due to blocked lines of sight or uncompensated lines of sight in average are not contemplated in this method.

In this method the errors are suposed to be gaussian in flux, not in magnitudes. The flux-averaging method minimizes the $\chi^2$ in a model-dependent
way. Here, we present how to flux-average for uncorrelated errors. First, one needs to convert the observed magnitude into ``fluxes'' $F(z_j)$

\begin{equation}
F(z_j) \equiv 10^{-(\mu_0(z_j)-25)/2.5} =  
\left( \frac{D_L^{data}(z)} {\mbox{Mpc}} \right)^{-2},
\end{equation}
where $\mu_0$ is the observed magnitude and $D_L^{data}$ is the measured distance.

The next step is to remove the redshift dependence, for a set of cosmological parameters $\{ {\bf s} \}$, by defining 
the ``absolute luminosities'' \{${\cal L}(z_j)$\}

\begin{equation}
{\cal L}(z_j) \equiv D_L^2(z_j |{\bf s})\,F(z_j).
\end{equation}

For each redshift bin $i$, the flux-average for the absolute luminosities $\left\{\overline{\cal L}^i\right\}$ in the mean 
redshift $\overline{z_i}$ is

\begin{equation} 
 \overline{\cal L}^i = \frac{1}{N}
 \sum_{j=1}^{N} {\cal L}^i_j(z^i_j),
 \hskip 1cm
 \overline{z_i} = \frac{1}{N}
 \sum_{j=1}^{N} z^i_j, 
\end{equation}
where $N$ is the number of bins.

Therefore, the binned flux in the $i$-th redshift bin is

\begin{equation}
\overline{F}(\overline{z}_i) = \overline{\cal L}^i /
d_L^2(\overline{z}_i|\mbox{\bf s}).
\end{equation}
The statistical analysis is done now with

\begin{equation}
\label{eq:chi2_flux}
\chi^2 = \sum_i \frac{ \left[\overline{F}(\overline{z}_i) -
F^p(\overline{z}_i|\mbox{\bf s}) \right]^2}{\sigma_{F,i}^2},
\end{equation}
where the errors $\sigma_{F,i}$ are the root mean square of the unbinned errors divided by the square root of the number of points in each bin.

\subsection{The modified DR approximation}
\label{mdr}

One criticism to the standard DR approach is that the change in the focusing due to less matter inside the beam must be compensated to curvature
and shear terms in order to provide the same local expansion rate as in the homogeneous case \cite{clarkson12}, what is not expected physically.
Thus, one way to improve the modelling is to allow a different expansion rate along the line of sight.

There are several possibilities to implement that, but for simplicity we use only one function, a modified smoothness parameter $\alpha_{\rm m}$,
to describe the effect. We write the Hubble parameter along the line of sight as

\begin{equation}
\frac{H(z)}{H_0}=[\alpha_{\rm m}(z)\Omega_{\rm m}(1+z)^3 + 1 - \Omega_{\rm m} +\Omega_{\rm m}(1-\alpha_{\rm m}(z))(1+z)^2] ^\frac{1}{2},  
\label{HubblezFRW2}
\end{equation}
where an extra curvature term appears because light propagates in an underdense medium which expands faster. We consider a constant smoothness
parameter which provides the highest difference from the standard flat $\Lambda$CDM case. For different approaches see \cite{mattsson,clarkson12}.
Note that Eq. (\ref{HubblezFRW2}) is considered only for the SNe light beams propagating on average in an underdense medium. Observations which infer the averaged $H(z)$,
as it is the case for the cosmic chronometers data, must be compared to the usual homogeneous Hubble parameter.  

\begin{figure*}[ht!]
     \begin{center}
        \subfigure[\, Union2.1 SNe Ia and $H(z)$ $(H_0(CR))$.]{%
            \label{fig:first_fig1}
            \includegraphics[width=0.30\textwidth]{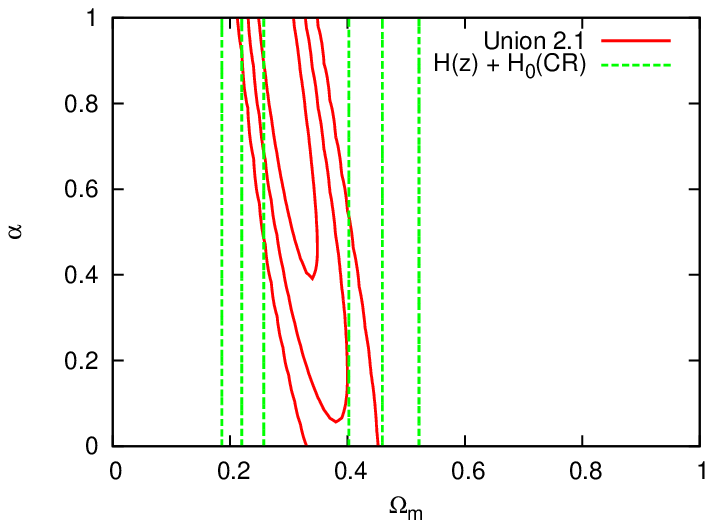}
        }%
        \subfigure[\, Union2.1 SNe Ia $+$ $H(z)$ $(H_0(CR))$.]{%
           \label{fig:second_fig1}
           \includegraphics[width=0.30\textwidth]{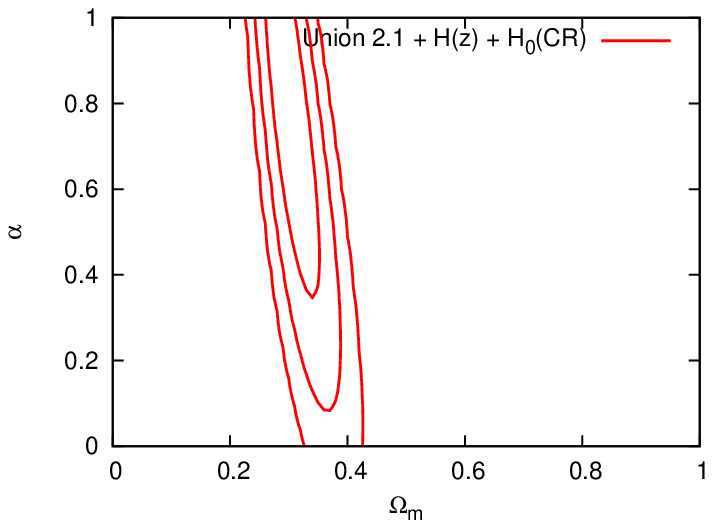}
        }
        \subfigure[\, Posterior probability for $\Omega_{\rm m}$.]{%
            \label{fig:third_fig1}
            \includegraphics[clip, width=0.30\textwidth]{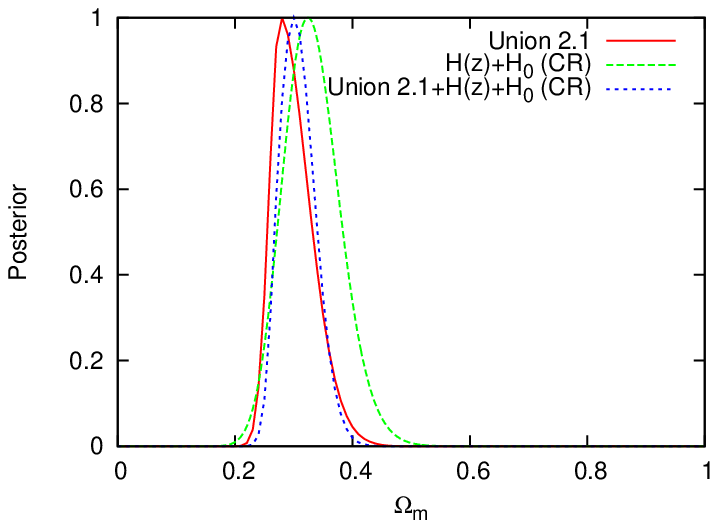}
        }\\
        \subfigure[\, Union2.1 SNe Ia and $H(z)$ $(H_0(R))$.]{%
            \label{fig:first_fig2}
            \includegraphics[width=0.3\textwidth]{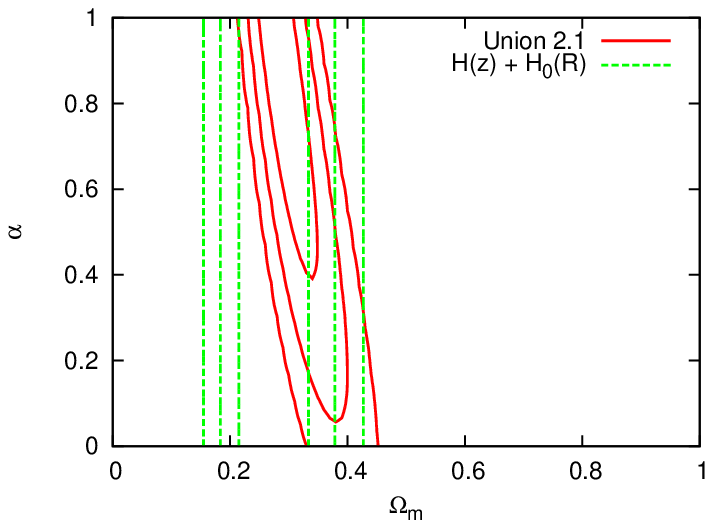}
        }%
        \subfigure[\, Union2.1 SNe Ia $+$ $H(z)$ $(H_0(R))$.]{%
           \label{fig:second_fig2}
           \includegraphics[width=0.3\textwidth]{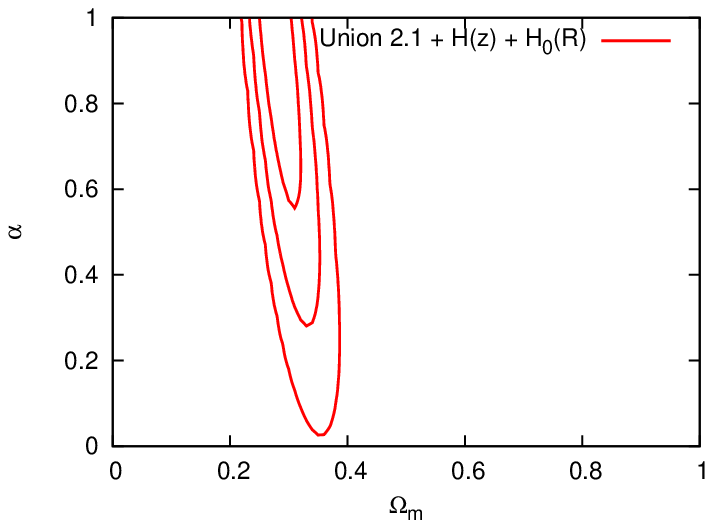}
        }
        \subfigure[\, Posterior probability for $\Omega_{\rm m}$.]{%
            \label{fig:third_fig2}
            \includegraphics[clip,width=0.3\textwidth]{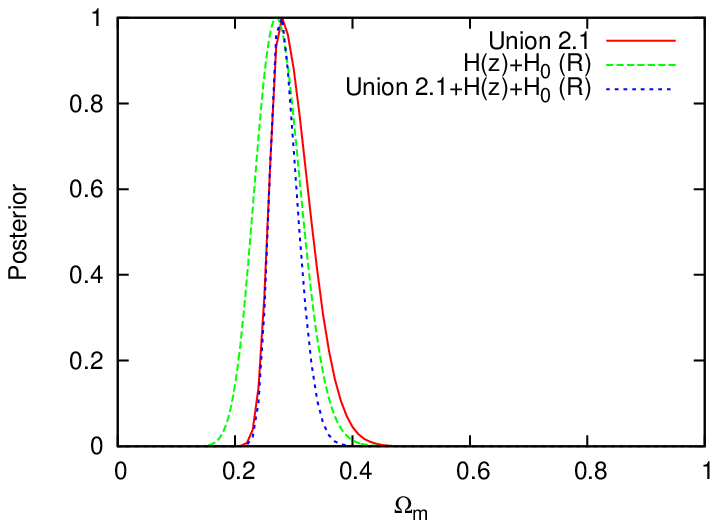}
        } 
        \subfigure[\, SDSS SNe Ia and $H(z)$ $(H_0(CR))$.]{%
            \label{fig:first_fig3}
            \includegraphics[width=0.3\textwidth]{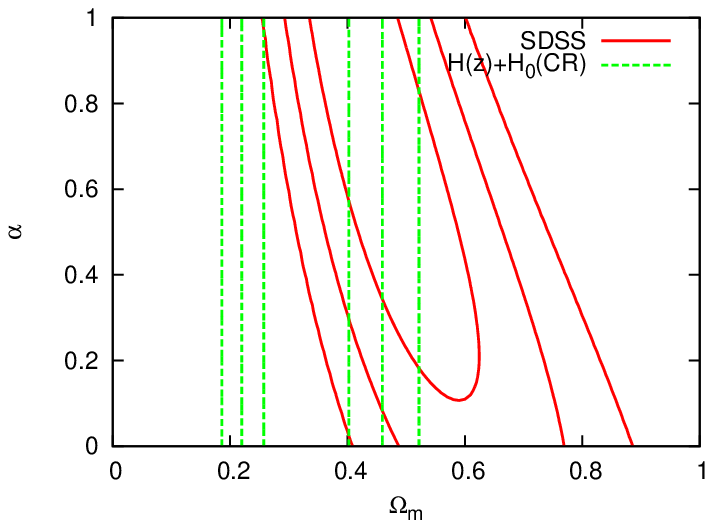}
        }%
        \subfigure[\, SDSS SNe Ia $+$ $H(z)$ $(H_0(CR))$.]{%
           \label{fig:second_fig3}
           \includegraphics[width=0.3\textwidth]{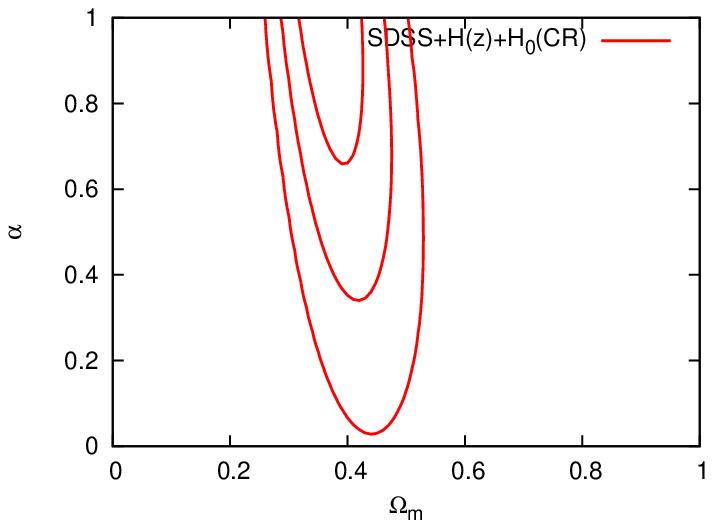}
        }
        \subfigure[\, Posterior probability for $\Omega_{\rm m}$.]{%
            \label{fig:third_fig3}
            \includegraphics[width=0.3\textwidth]{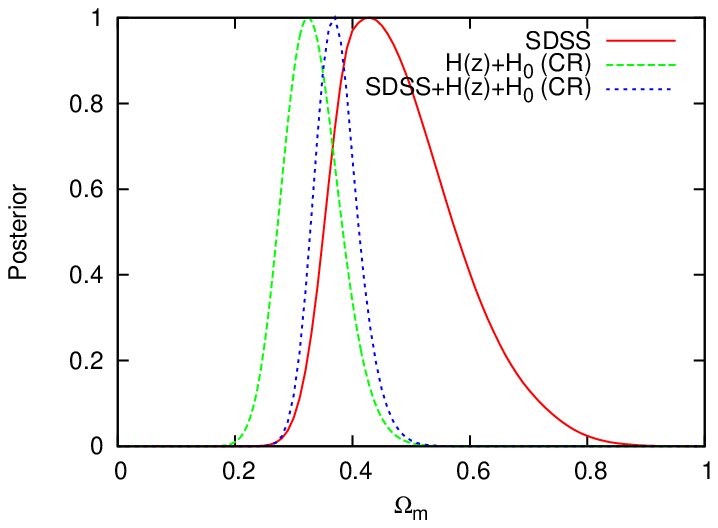}
        }\\
        \subfigure[\, SDSS SNe Ia and $H(z)$ $(H_0(R))$.]{%
            \label{fig:first_fig4}
            \includegraphics[width=0.3\textwidth]{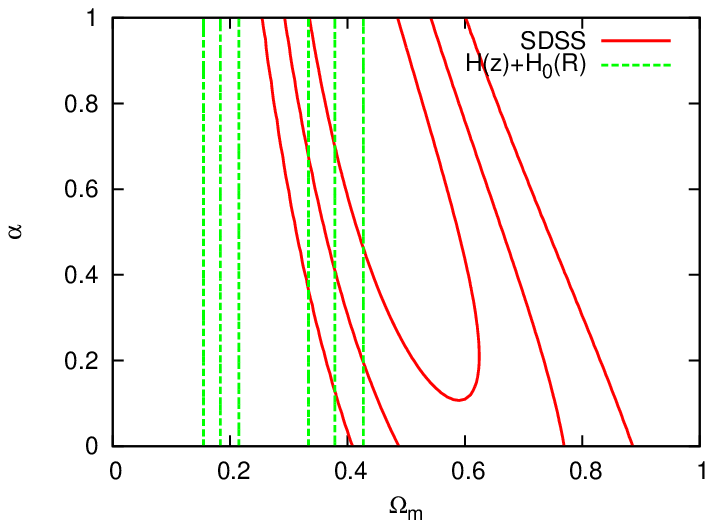}
        }%
        \subfigure[\, SDSS SNe Ia $+$ $H(z)$ $(H_0(R))$.]{%
           \label{fig:second_fig4}
           \includegraphics[width=0.3\textwidth]{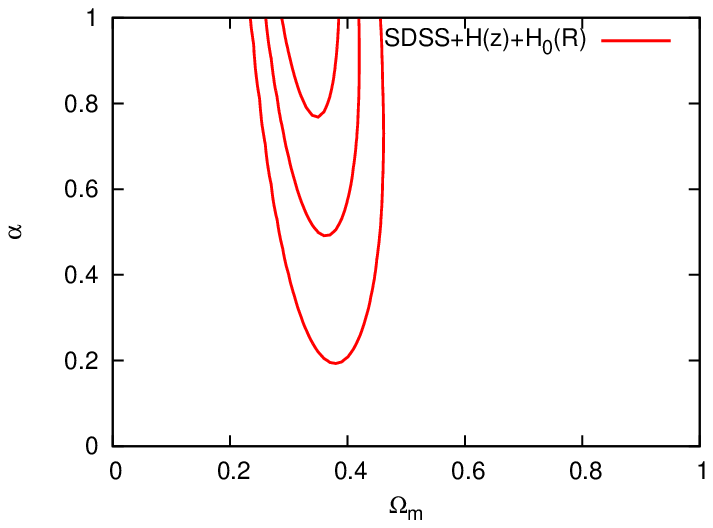}
        }
        \subfigure[\, Posterior probability for $\Omega_{\rm m}$.]{%
            \label{fig:third_fig4}
            \includegraphics[width=0.3\textwidth]{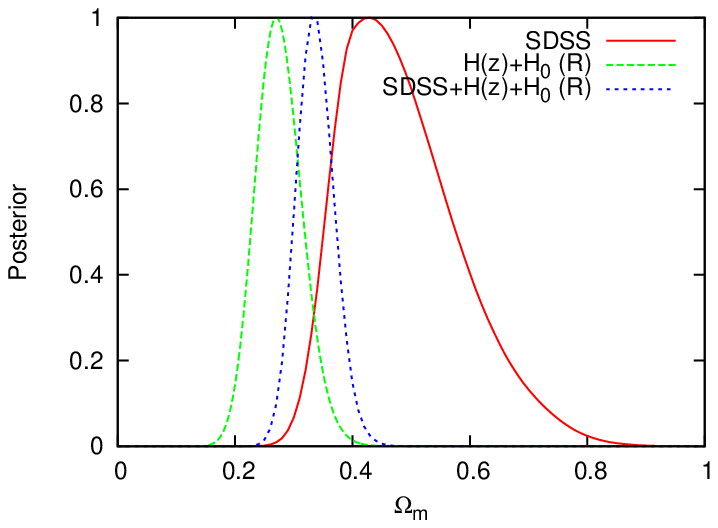}
        }

    \end{center}
    \caption{\emph{The Dyer-Roeder approximation}. The left panels refer to the $(\Omega_{\rm m},\alpha)$ plane for a flat $\Lambda$CDM model with SNe Ia data (red solid contours) and
    $H(z)$ measurements. The contours represent 
    the 68.3\%, 95.4\%, and 99.7\% confidence levels. The middle panels refer to  
    a joint analysis involving the two samples. The right panels refer to the posterior probability for $\Omega_{\rm m}$. The red solid line stands for the SNe Ia posterior,
    the green dashed line for the $H(z)$ posterior, and the blue dotted line for the joint posterior.  
     }%
   \label{fig1}
\end{figure*}

\section{Samples and Results}
\label{sec_res}

\begin{figure*}[ht!]
     \begin{center}
        \subfigure[\, Union2.1 SNe Ia and $H(z)$ $(H_0(CR))$.]{%
            \label{fig:first_fig5}
            \includegraphics[width=0.3\textwidth]{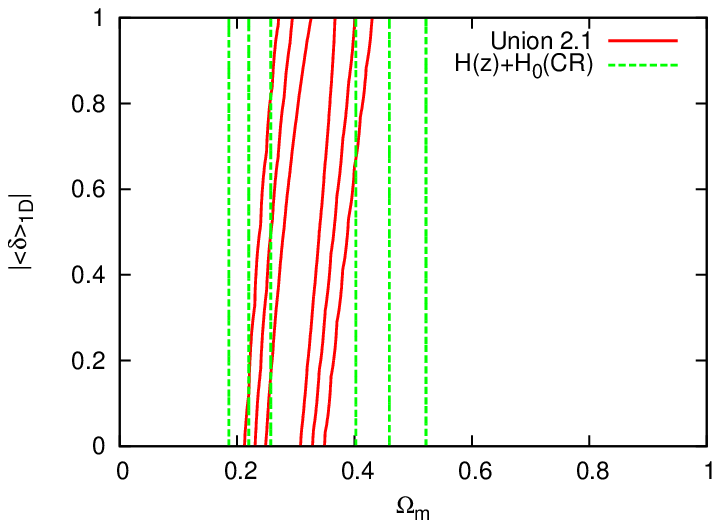}
        }%
        \subfigure[\, Union2.1 SNe Ia $+$ $H(z)$ $(H_0(CR))$.]{%
           \label{fig:second_fig5}
           \includegraphics[width=0.3\textwidth]{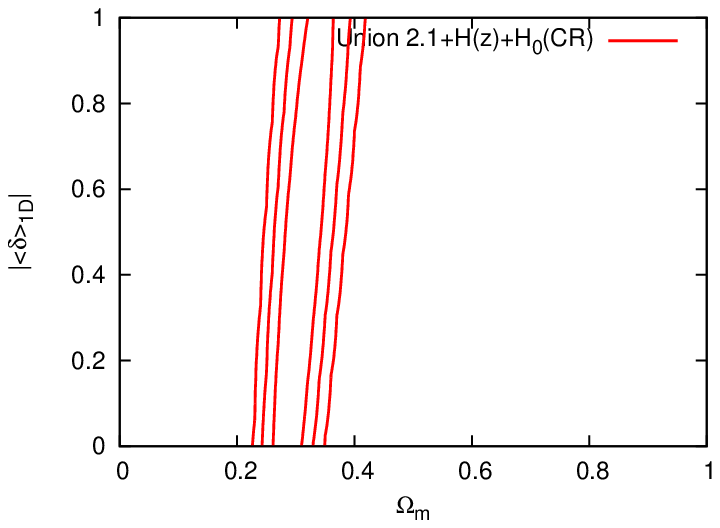}
        }
        \subfigure[\, Posterior probability for $\Omega_{\rm m}$.]{%
            \label{fig:third_fig5}
            \includegraphics[width=0.3\textwidth]{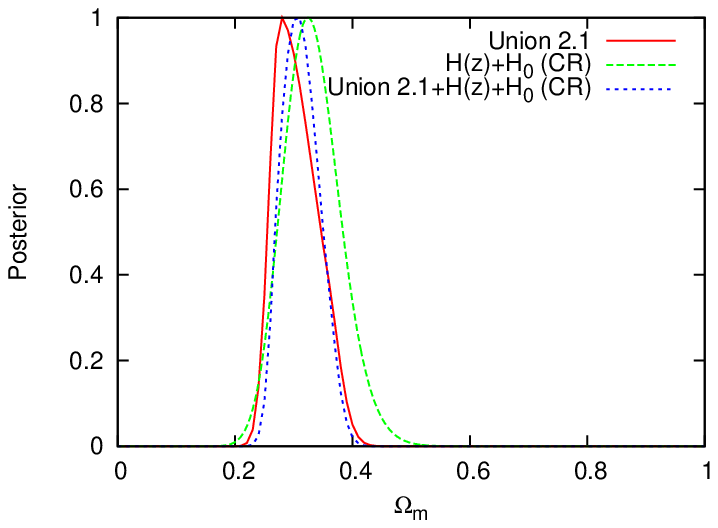}
        }\\
        \subfigure[\, Union2.1 SNe Ia and $H(z)$ $(H_0(R))$.]{%
            \label{fig:first_fig6}
            \includegraphics[width=0.3\textwidth]{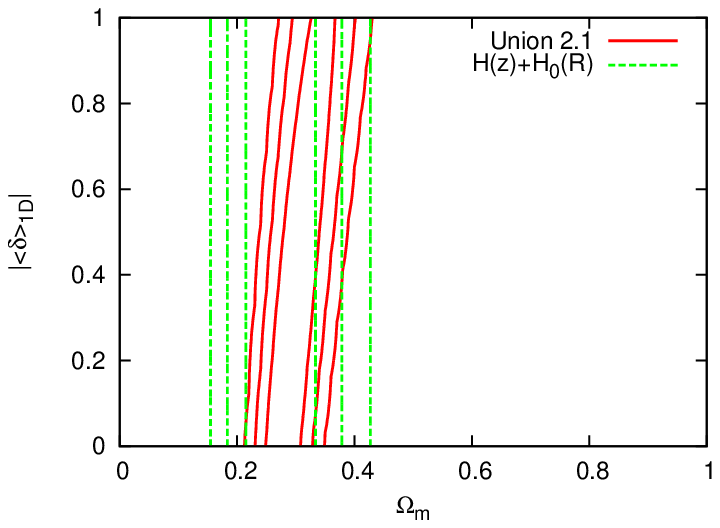}
        }%
        \subfigure[\, Union2.1 SNe Ia $+$ $H(z)$ $(H_0(R))$.]{%
           \label{fig:second_fig6}
           \includegraphics[width=0.3\textwidth]{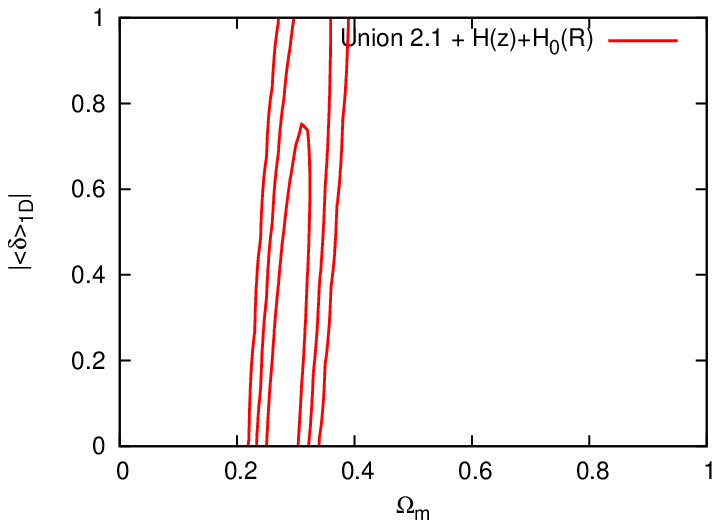}
        }
        \subfigure[\, Posterior probability for $\Omega_{\rm m}$.]{%
            \label{fig:third}
            \includegraphics[width=0.3\textwidth]{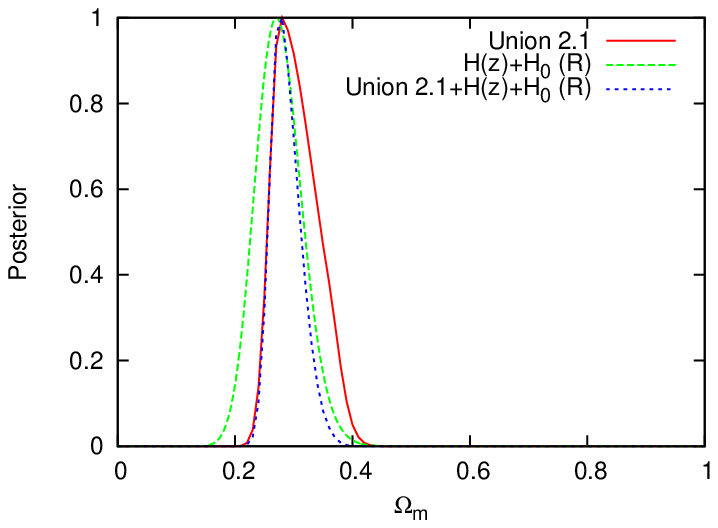}
        }\\
        \subfigure[\, SDSS SNe Ia and $H(z)$ $(H_0(CR))$.]{%
            \label{fig:first_fig7}
            \includegraphics[width=0.3\textwidth]{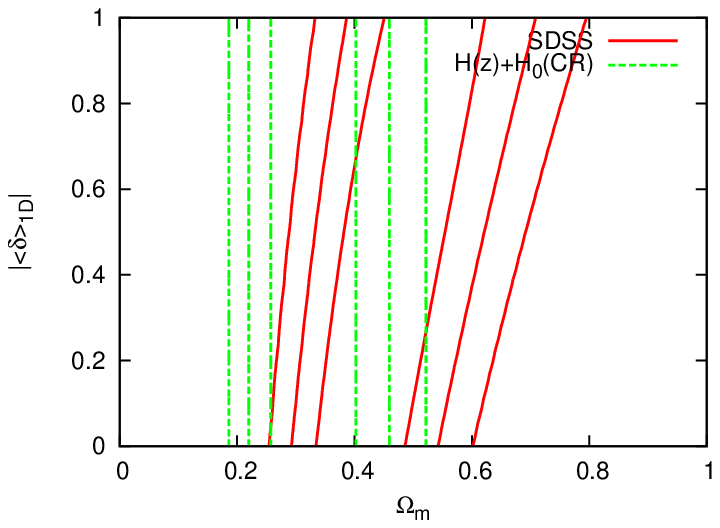}
        }%
        \subfigure[\, SDSS SNe Ia $+$ $H(z)$ $(H_0(CR))$.]{%
           \label{fig:second_fig7}
           \includegraphics[width=0.3\textwidth]{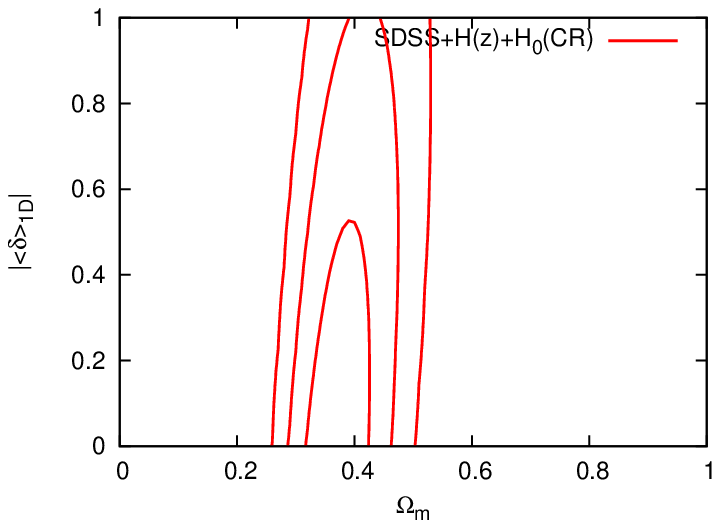}
        }
        \subfigure[\, Posterior probability for $\Omega_{\rm m}$.]{%
            \label{fig:third_fig7}
            \includegraphics[width=0.3\textwidth]{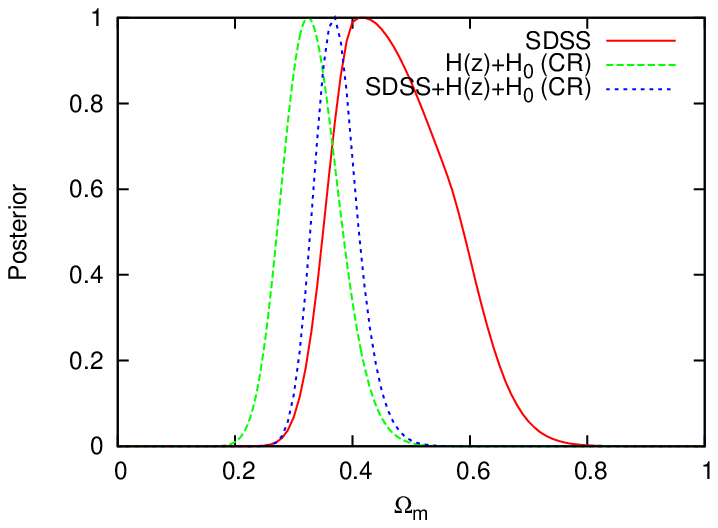}
        }\\
        \subfigure[\, SDSS SNe Ia and $H(z)$ $(H_0(R))$.]{%
            \label{fig:first_fig8}
            \includegraphics[width=0.3\textwidth]{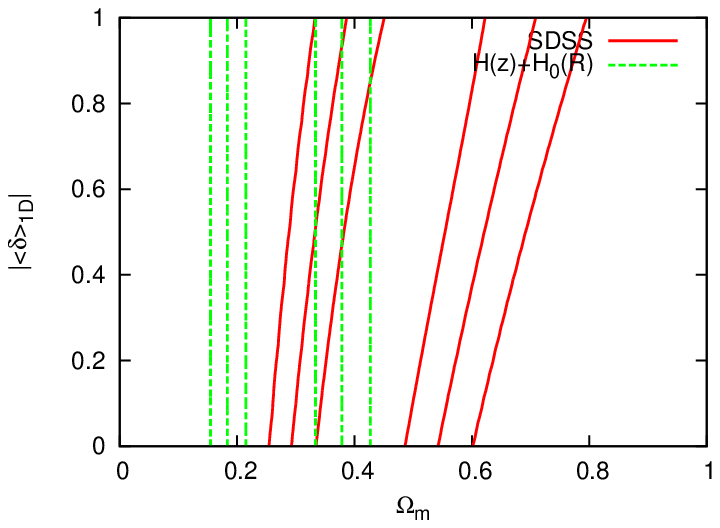}
        }%
        \subfigure[\, SDSS SNe Ia $+$ $H(z)$ $(H_0(R))$.]{%
           \label{fig:second_fig8}
           \includegraphics[width=0.3\textwidth]{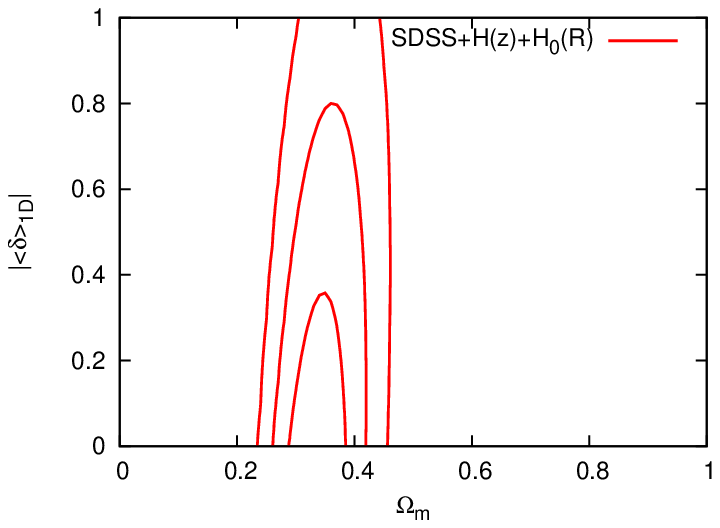}
        }
        \subfigure[\, Posterior probability for $\Omega_{\rm m}$.]{%
            \label{fig:third_fig8}
            \includegraphics[width=0.3\textwidth]{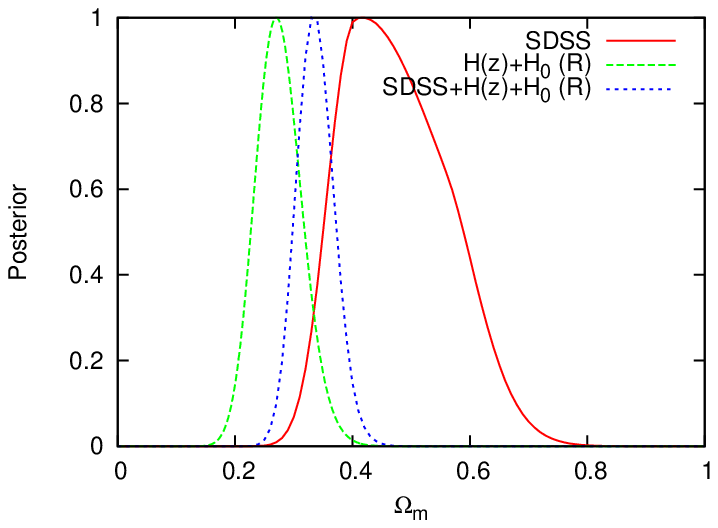}
        }
    \end{center}
    \caption{\emph{The weak-lensing approximation}. The left panels refer to the $(\Omega_{\rm m},|\langle\delta\rangle_{1D}|)$ plane for a flat $\Lambda$CDM model with SNe Ia data (red solid      contours) and $H(z)$ measurements. The contours represent  the 68.3\%, 95.4\%, and 99.7\% confidence levels. The middle panels refer to  
    a joint analysis involving the two samples. The right panels refer to the posterior probability for $\Omega_{\rm m}$. The red solid line stands for the SNe Ia posterior,
    the green dashed line for the $H(z)$ posterior, and the blue dotted line for the joint posterior. 
     }%
   \label{fig2}
\end{figure*}

In this section we perform statistical analyses to constrain the parameters of the models described above. In order to have a broad coverage
of different observational probes, two samples of SNe Ia are considered calibrated with different light curve fitters. The first is constituted
of 581 SNe Ia from \cite{union2.1,rubin}, the Union2.1 compilation data calibrated with SALT2 \cite{salt2}. The second sample has 288 SNe Ia calibrated
with the MLCS2k2 fitter \cite{mlcs} called the SDSS sample.

Except for the flux-averaging scheme, the set of parameters ${\bf p}$ are constrained by maximizing the posterior probability, which is equivalent,
for gaussianly distributed measurements, to minimize the $\chi^2$ function
\begin{equation}
 \chi^2 = \sum_i \left[ \frac{\mu_{th}({\bf p},z_i) - \mu_{0}(z_i)}{\sigma_{\mu_0}(z_i)} \right]^2,
\end{equation}
where $\mu_{th}=5 \log(D_L/{Mpc})+25$ is the theoretical distance modulus, $\mu_{0}$ is the observed distance modulus and $\sigma_{\mu_0}$ its 
respective uncertainty. We treat $H_0$ as a nuisance parameter and we marginalize over it.

We also consider 19 $H(z)$ measurements from differential age of 
passively evolving galaxies, also called the cosmic chronometers data \cite{hzmeasurements}. This sample was used to constrain several cosmological parameters in \cite{moresco}, 
and in addition with
$H(z)$ from BAO measurements to investigate dark energy models \cite{farooq}. We do not use the $H(z)$ from BAO measurements since it is not clear how the inhomogeneities can 
affect such measurements. Although expected to be small due to the inference from large scales where the effects may average out, 
we decided to be more conservative. The statistical analyses follow as the same for SNe Ia, with the $\chi^2$ given by

\begin{equation}
 \chi^2 = \sum_i^{19} \left[ \frac{H_{th}({\bf p},z_i) - H_{obs}(z_i)}{\sigma_{H_{obs},i}} \right]^2,
\end{equation}
where $H_{th}$ is the theoretical Hubble parameter, $H_{obs}$ is the observed Hubble parameter, and $\sigma_{H_{obs}}$ its respective uncertainty.
Again, we marginalize over $H_0$ with a gaussian prior following Ref. \cite{farooq}.

In order to deal with a tension between measurements of $H_0$, two values are used in the analyses. The first is based on {\it Hubble
Space Telescope} measurements: $\bar H_0(R) = 73.8 \pm 2.4$ km s$^{-1}$ Mpc$^{-1}$ \cite{H0riess}. The second is based on a median statistics 
analysis of 553 values for $H_0$: $\bar H_0 (CR)= 68.0 \pm 2.8$ km s$^{-1}$ Mpc$^{-1}$ \cite{H0CR}. The last value 
is in agreement with the latest measurement from {\it PLANCK} \cite{planck}.

\subsection{The DR approximation} 
\label{DR} 

In Fig. \ref{fig1}(a)-(c) we display the results of the statistical analyses considering 581 from the Union2.1 compilation data \cite{union2.1,rubin}, 19 
$H(z)$ measurements \cite{hzmeasurements} and $H_0$(CR) \cite{H0CR} for a flat $\Lambda$CDM model. It is shown that the smoothness parameter 
is weakly constrained with SNe Ia data
and the addition of the $H(z)$ measurements do not improve the limits due to the degeneracy of the data in the parameter space. For the SNe Ia data
only, the parameters are constrained to be in the intervals within the 95.4\% confidence level $(2\sigma)$: 
$\alpha=0.98^{+0.02}_{-0.76}$ and $\Omega_{\rm m}=0.28^{+0.08}_{-0.04}$.
As expected, the $H(z)$ can only improve the constraints through a joint analysis \cite{bs2011}, since the expansion rate in the DR approximation 
is the same of a homogeneous model. The restrictions obtained in the joint analysis are: $\alpha=0.81^{+0.19}_{-0.59}$ and 
$\Omega_{\rm m}=0.30^{+0.06}_{-0.04}$ $(2\sigma)$.

In Fig. \ref{fig1}(d)-(f) it is shown the results when a higher value for $H_0$(R) \cite{H0riess} is considered. In this case the $H(z)$ constraints are 
slightly shifted to the left allowing better constraints to $\alpha$ in the joint analysis, which provides
$\alpha=0.98^{+0.02}_{-0.55}$ and $\Omega_{\rm m}=0.28^{+0.05}_{-0.03}$ $(2\sigma)$.

At this point it is interesting to compare these results with previous analyses. Our results are fully compatible with constraints based on analyses
involving SNe Ia, compact radio sources, gamma-ray bursts and $H(z)$ measurements \cite{bs2011,sl2008,scl2008,bsl2012,yang2013}.
The same trend noticed in \cite{bsl2012} remains, where a larger sample of SNe Ia weakens the constraints over the smoothness parameter.
Shortly, the DR approximation can handle with all observations so far. 

On the other hand, the results are not in
agreement with a recent analysis involving SNe Ia, gamma-ray bursts and $H(z)$ measurements \cite{breton2013}. So, it is necessary to understand
the difference between their results and ours. First of all, our SNe Ia constraints are very different, although we added only one high-redshift
supernova. We ascribe this difference possibly by a lack of convergence in their MCMC analysis, which is in general problematic when the best fit is
near the borders of the considered interval. Note that $\alpha=1$ is only allowed at $3\sigma$ (their table III), which is the cosmic concordance model.
Second, as emphasized in \cite{bs2011}, the $H(z)$ measurements are completely independent of $\alpha$, since in the DR approximation light propagates
in a universe with local density $\alpha \rho_m$, but the same expansion rate of a perfectly homogeneous universe. Thus, their equation (22) is not 
valid. Actually, it is the breakdown of this equation which turns possible to use consistency tests \cite{consistency} to see whether 
the effects of the inhomogeneities in light propagation are significant \cite{bl2012}. To finish, although the use of gamma-ray bursts may give
a lever arm in the Hubble diagram to discern among cosmological models, caution is needed, since the understanding of the phenomenological 
relations used to calibrate them is still incipient.

\begin{table}[htbp]
\caption{Limits to $\alpha$ and $\Omega_{\rm m}$ in the DR approximation.}
\label{table1}
\begin{center}
\begin{tabular}{@{}cccc@{}}
\hline Sample & $\Omega_{\rm m}$ ($2\sigma$) & $\alpha$ ($2\sigma$) &
$\chi^2_{min}$
\\ \hline\hline
Union2.1 & $0.28^{+0.08}_{-0.04}$ & $0.98^{+0.02}_{-0.76}$ & $562.2$ \\
Union2.1 $+$ $H(z)$ $+$ $H_0$(CR)  & $0.30^{+0.06}_{-0.04}$  & $0.81^{+0.19}_{-0.59}$ & $576.5$ \\
Union2.1 $+$ $H(z)$ $+$ $H_0$(R) & $0.28^{+0.05}_{-0.03}$ & $0.98^{+0.02}_{-0.55}$ & $577.2$ \\
SDSS & $0.43^{+0.26}_{-0.11}$ & unconstrained & $154.8$ \\
SDSS $+$ $H(z)$ $+$ $H_0$(CR)  &$0.37^{+0.07}_{-0.06}$  & $1.0^{+0.0}_{-0.49}$     &$169.9$ \\
SDSS $+$ $H(z)$ $+$ $H_0$(R) & $0.33^{+0.06}_{-0.05}$ & $1.0^{+0.0}_{-0.36}$ & $174.4$ \\
\hline
\end{tabular}
\end{center}
\end{table}

In order to see how different fitters affect the constraints, we consider the 288 SNe Ia from the SDSS compilation data \cite{sdss} calibrated with the 
MLCS2k2 fitter \cite{mlcs}. This sample was chosen because these data prefer more exotic models compared to the standard flat
$\Lambda$CDM model \cite{exotic}. So, a natural question arises if the inhomogeneities may give a better agreement with other cosmic probes. In this
sense, as we did for the Union2.1 compilation data, a comparison is made between the constraints derived from SNe Ia and those from the 
$H(z)$ measurements, again for the two values of $H_0$.

Figure \ref{fig1}(g)-(i) shows the results of the statistical analyses involving the 288 SNe Ia from the SDSS compilation data \cite{sdss}, 19 $H(z)$ 
measurements \cite{hzmeasurements} and $H_0$(CR) \cite{H0CR}. There is a small region in the paramter space where the data agree at $1\sigma$
shown in Fig. \ref{fig:first_fig3}. The different degenerescence in the parameter space allows better constraints to $\alpha$ in the joint
analysis, where the parameters are restricted to the intervals:
$\alpha=1.0^{+0.0}_{-0.49}$ and $\Omega_{\rm m}=0.37^{+0.07}_{-0.06}$ $(2\sigma)$.

\begin{figure*}[ht!]
     \begin{center}
        \subfigure[\, Posterior probability for $\Omega_{\rm m}$.]{%
            \label{fig:first_fig9}
            \includegraphics[width=0.4\textwidth]{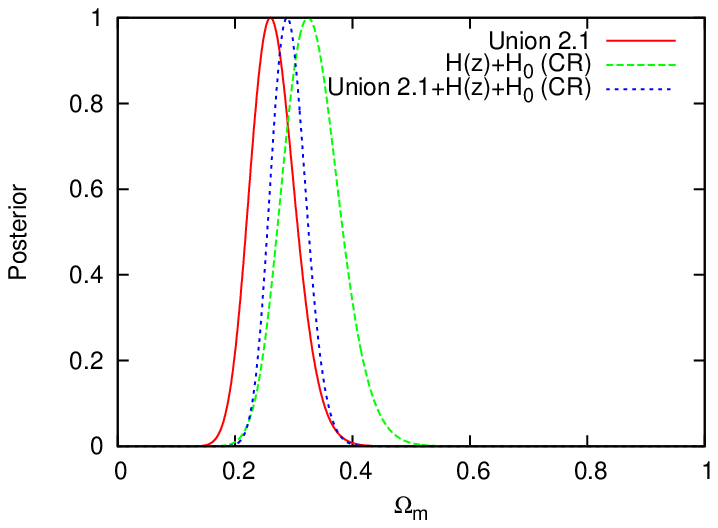}
        }%
        \subfigure[\, Posterior probability for $\Omega_{\rm m}$.]{%
           \label{fig:second_fig9}
           \includegraphics[width=0.4\textwidth]{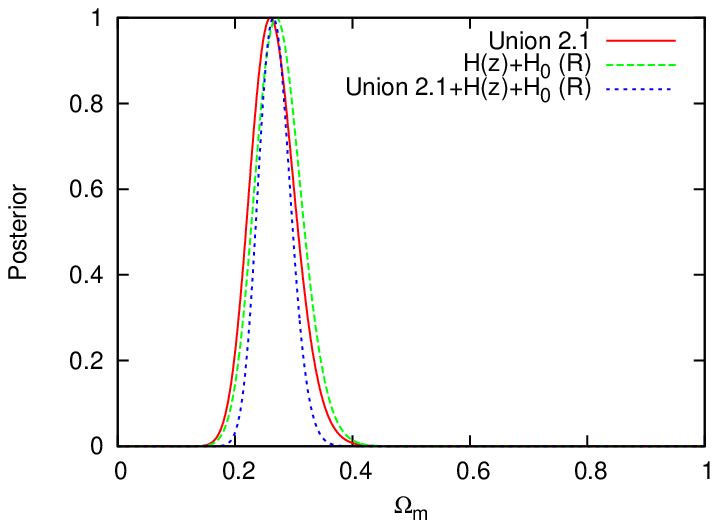}
        }\\
        \subfigure[\, Posterior probability for $\Omega_{\rm m}$.]{%
            \label{fig:first_fig10}
            \includegraphics[width=0.4\textwidth]{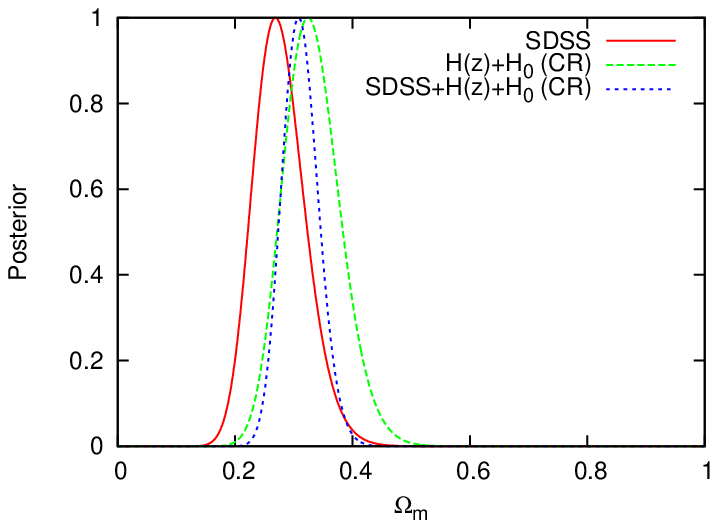}
        }%
        \subfigure[\, Posterior probability for $\Omega_{\rm m}$.]{%
           \label{fig:second_fig10}
           \includegraphics[width=0.4\textwidth]{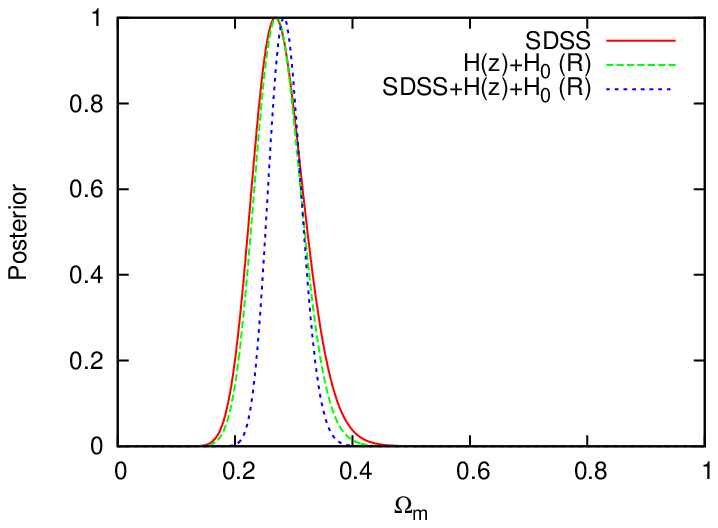}
        }

    \end{center}
    \caption{%
        \emph{The flux-averaging approximation}. Posterior probability for $\Omega_{\rm m}$. (a) The results for the Union2.1 compilation data \cite{union2.1,rubin} are 
        displayed by the solid red line. The dashed green line stands for the analysis considering 19 $H(z)$ measurements \cite{hzmeasurements} and
        $H_0$(CR) \cite{H0CR}. The dotted blue line refers to the joint analysis. (b) The same as (a) with $H_0$(R) \cite{H0riess}.
        Panels (c) and (d) show the results of the same analyses as (a) and (b), but now considering the SDSS compilation data \cite{sdss}. 
     }%
   \label{fig3}
\end{figure*}

Figure \ref{fig1}(j)-(l) displays the results with the same samples of Figs. \ref{fig1}(g)-(i), but with $H_0$(R) \cite{H0riess}. As one can see in Fig. \ref{fig:first_fig4}, the 
samples are in tension at $1\sigma$, which implies that a higher value for $H_0$ cannot alleviate the tension between the SNe Ia sample (with
inhomogeneities), and other cosmic probes in a flat $\Lambda$CDM model. For the joint analysis, the parameters are constrained to be:
$\alpha=1.0^{+0.0}_{-0.36}$ and $\Omega_{\rm m}=0.33^{+0.06}_{-0.05}$ $(2\sigma)$. These results show the importance of accurate $H_0$ measurements
and their impact on other cosmological parameters. Table \ref{table1} summarizes the results obtained from the analyses of the DR approximation.

\begin{figure*}[ht!]
     \begin{center}
        \subfigure[\, Union2.1 SNe Ia and $H(z)$ $(H_0(CR))$.]{%
            \label{fig:first_fig11}
            \includegraphics[width=0.30\textwidth]{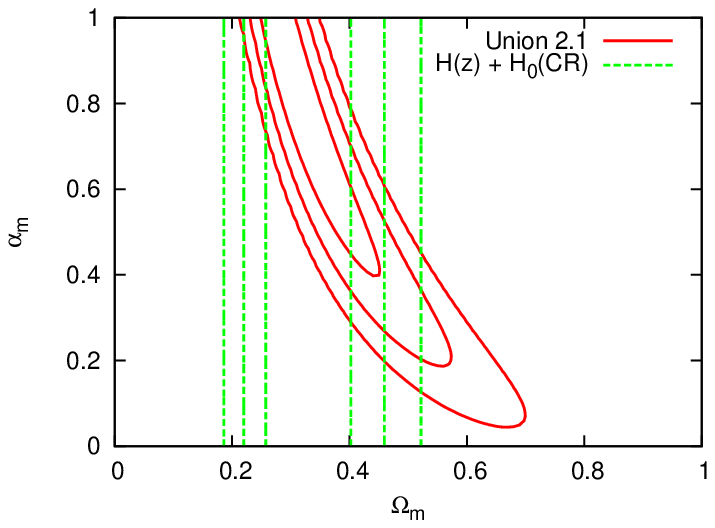}
        }%
        \subfigure[\, Union2.1 SNe Ia $+$ $H(z)$ $(H_0(CR))$.]{%
           \label{fig:second_fig11}
           \includegraphics[width=0.30\textwidth]{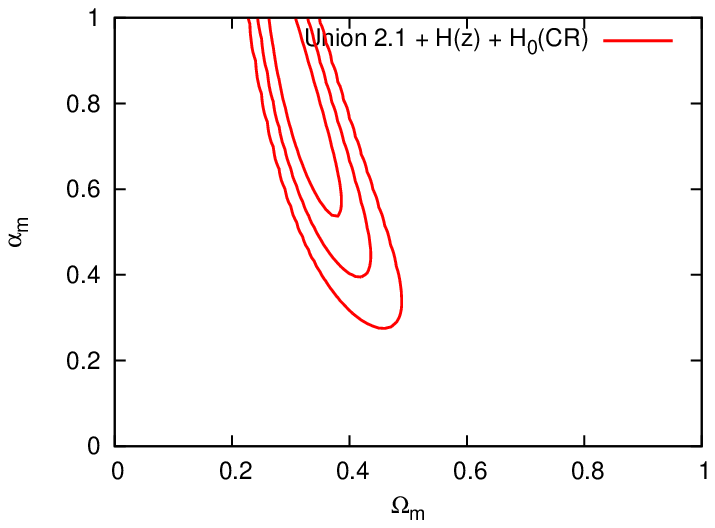}
        }
        \subfigure[\, Posterior probability for $\Omega_{\rm m}$.]{%
            \label{fig:third_fig11}
            \includegraphics[clip,width=0.30\textwidth]{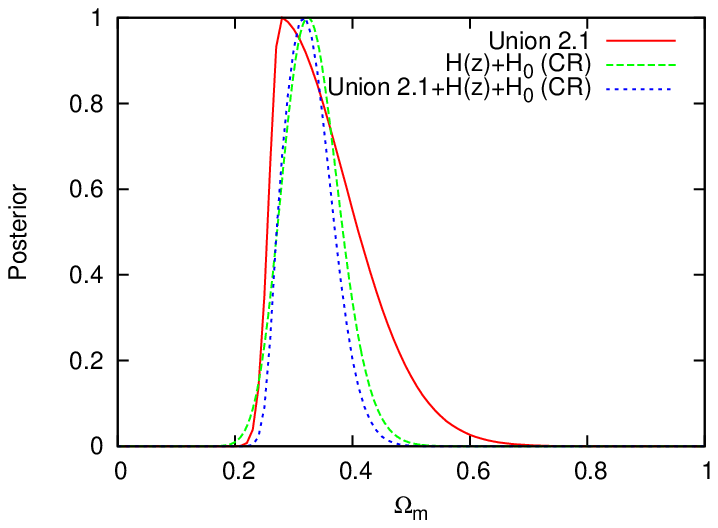}
        }\\
        \subfigure[\, Union2.1 SNe Ia and $H(z)$ $(H_0(R))$.]{%
            \label{fig:first_fig12}
            \includegraphics[width=0.3\textwidth]{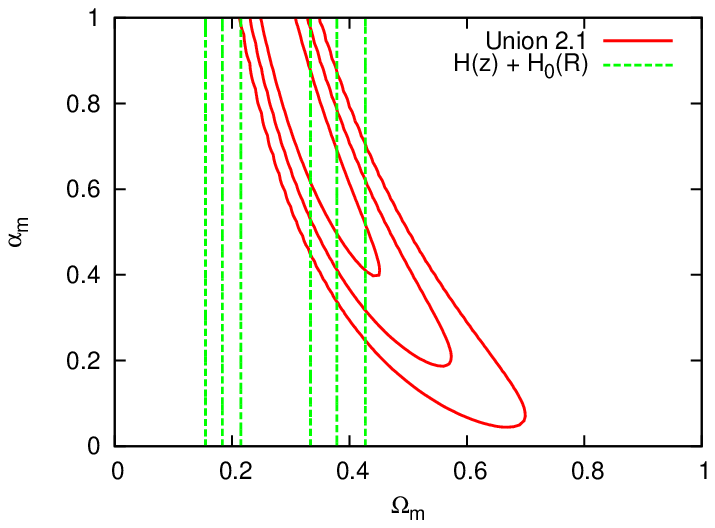}
        }%
        \subfigure[\, Union2.1 SNe Ia $+$ $H(z)$ $(H_0(R))$.]{%
           \label{fig:second_fig12}
           \includegraphics[width=0.3\textwidth]{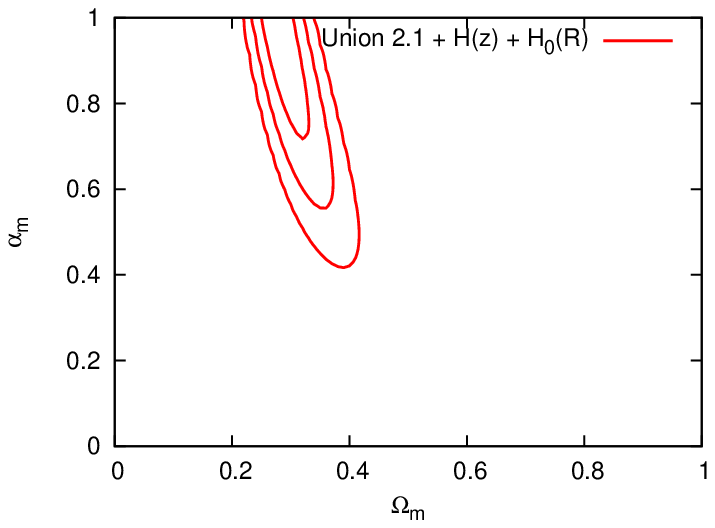}
        }
        \subfigure[\, Posterior probability for $\Omega_{\rm m}$.]{%
            \label{fig:third_fig12}
            \includegraphics[clip, width=0.3\textwidth]{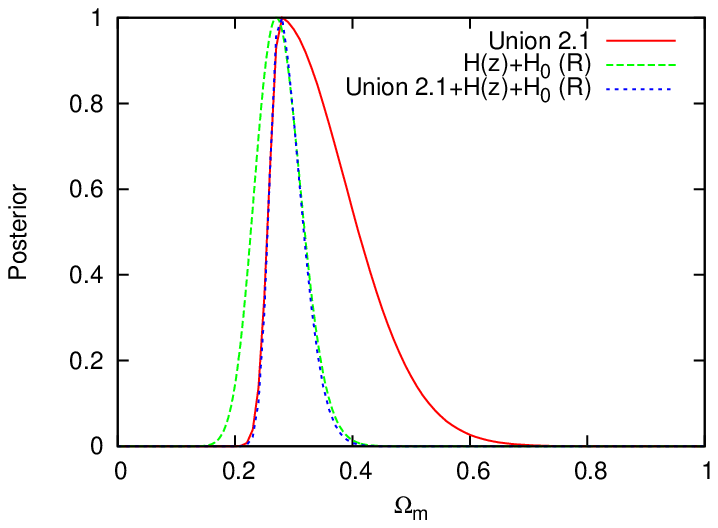}
        }\\
        \subfigure[\, SDSS SNe Ia and $H(z)$ $(H_0(CR))$.]{%
            \label{fig:first_fig13}
            \includegraphics[width=0.3\textwidth]{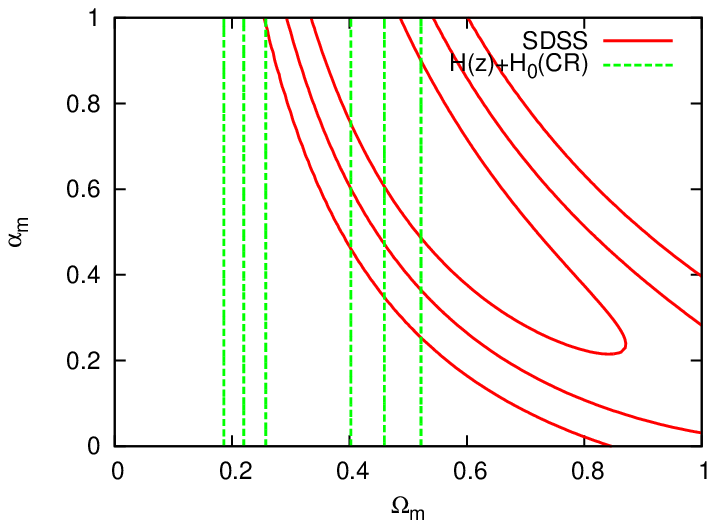}
        }%
        \subfigure[\, SDSS SNe Ia $+$ $H(z)$ $(H_0(CR))$.]{%
           \label{fig:second_fig13}
           \includegraphics[width=0.3\textwidth]{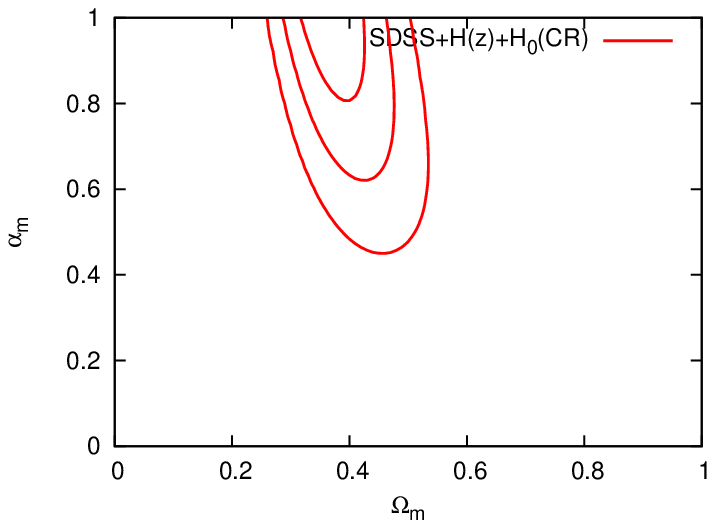}
        }
        \subfigure[\, Posterior probability for $\Omega_{\rm m}$.]{%
            \label{fig:third_fig13}
            \includegraphics[width=0.3\textwidth]{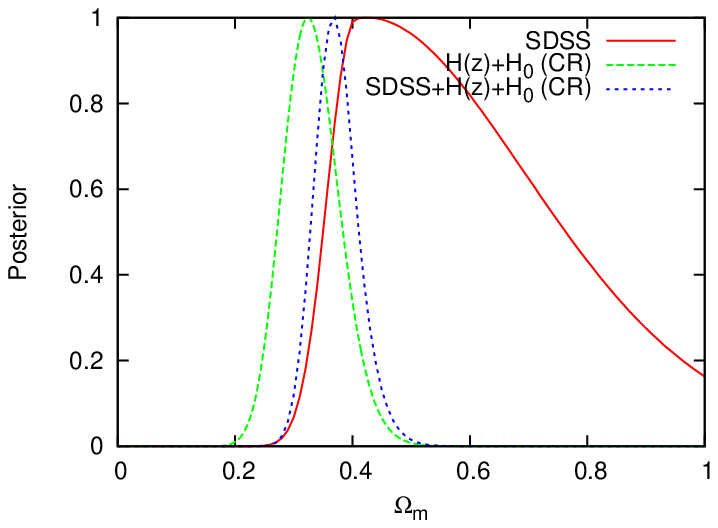}
        }\\ 
        \subfigure[\, SDSS SNe Ia and $H(z)$ $(H_0(R))$.]{%
            \label{fig:first_fig14}
            \includegraphics[width=0.3\textwidth]{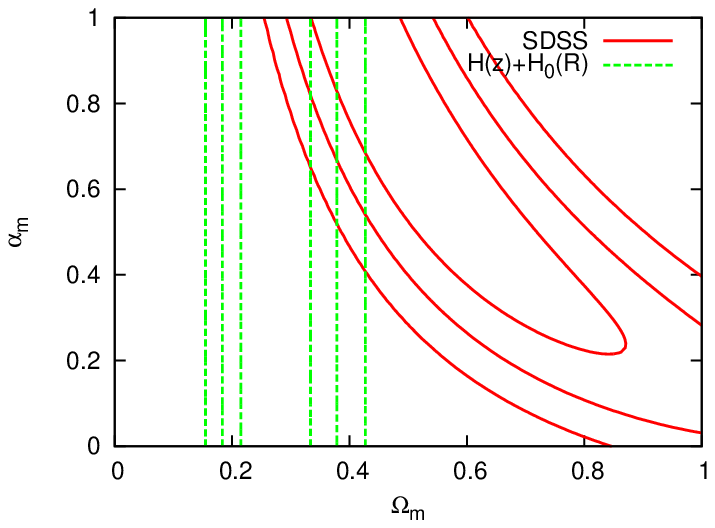}
        }%
        \subfigure[\, SDSS SNe Ia $+$ $H(z)$ $(H_0(R))$.]{%
           \label{fig:second_fig14}
           \includegraphics[width=0.3\textwidth]{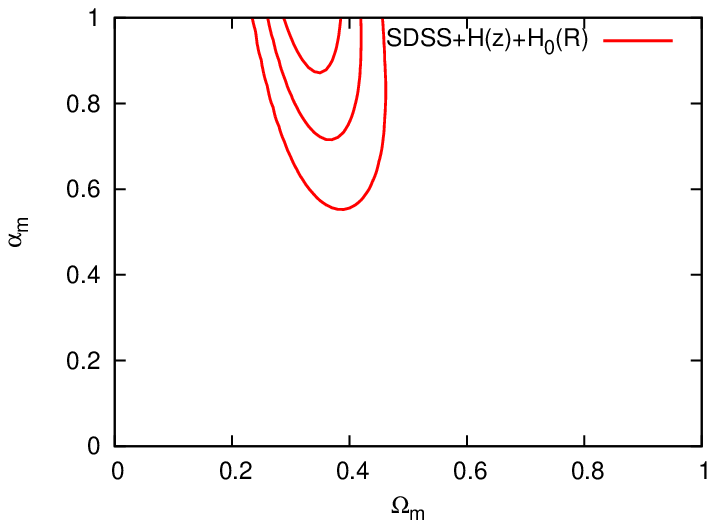}
        }
        \subfigure[\, Posterior probability for $\Omega_{\rm m}$.]{%
            \label{fig:third_fig14}
            \includegraphics[width=0.3\textwidth]{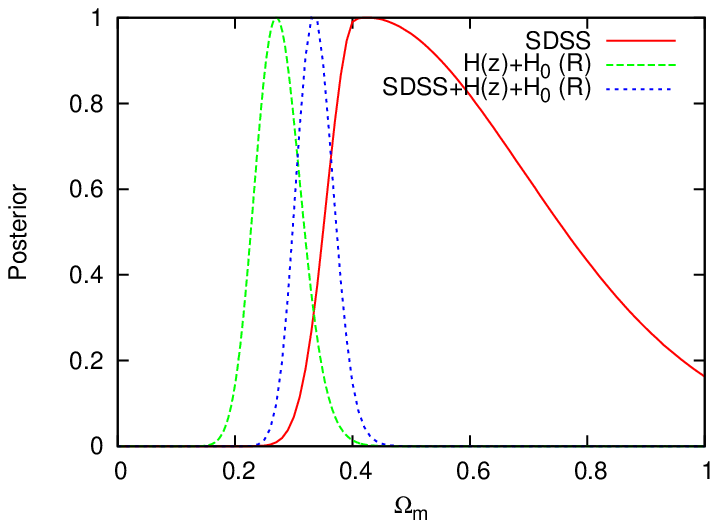}
        }
    \end{center}
    \caption{\emph{The modified Dyer-Roeder approximation}. The left panels refer to the $(\Omega_{\rm m},\alpha_{\rm m})$ plane for a flat $\Lambda$CDM model with SNe Ia data (red solid contours) and
    $H(z)$ measurements. The contours represent 
    the 68.3\%, 95.4\%, and 99.7\% confidence levels. The middle panels refer to  
    a joint analysis involving the two samples. The right panels refer to the posterior probability for $\Omega_{\rm m}$. The red solid line stands for the SNe Ia posterior,
    the green dashed line for the $H(z)$ posterior, and the blue dotted line for the joint posterior.    
     }%
   \label{fig4}
\end{figure*}

\begin{table}[htbp]
\caption{Limits to $|\langle\delta\rangle_{1D}|$ and $\Omega_{\rm m}$ in the weak lensing approximation.}
\label{table2}
\begin{center}
\begin{tabular}{@{}cccc@{}}
\hline Sample & $\Omega_{\rm m}$ ($2\sigma$) & $|\langle\delta\rangle_{1D}|$ ($2\sigma$) &
$\chi^2_{min}$
\\ \hline\hline
Union2.1 & $0.28^{+0.10}_{-0.04}$ & unconstrained & $562.2$ \\
Union2.1 $+$ $H(z)$ $+$ $H_0$(CR)  & $0.31^{+0.06}_{-0.05}$  & unconstrained & $576.4$ \\
Union2.1 $+$ $H(z)$ $+$ $H_0$(R) & $0.28^{+0.06}_{-0.03}$ & $0.04^{+0.95}_{-0.04}$ & $577.2$ \\
SDSS & $0.42^{+0.24}_{-0.10}$ & unconstrained & $154.8$ \\
SDSS $+$ $H(z)$ $+$ $H_0$(CR)  &$0.37^{+0.07}_{-0.06}$  & $0.0^{+0.78}_{-0.0}$     &$169.9$ \\
SDSS $+$ $H(z)$ $+$ $H_0$(R) & $0.33^{+0.06}_{-0.05}$ & $0.0^{+0.56}_{-0.0}$ & $174.4$ \\
\hline
\end{tabular}
\end{center}
\end{table}

\subsection{The weak lensing approximation} 
\label{BOL} 

In order to test the weak lensing approximation the relation between it and the DR approximation is used, where the smoothness parameter
obeys \cite{bol_dr_2011}

\begin{equation}
 \alpha(z) = 1 - \frac{|\langle\delta\rangle_{1D}|}{(1+z)^{5\over4}}.
\end{equation}
Note that we restrict $|\langle\delta\rangle_{1D}|$ in the interval $[0,1]$, which implies $\alpha$ is in the interval $[0,1]$ as well. Hence, light can 
propagate in an underdense medium in this approximation.

Figure \ref{fig2} shows the results of the statistical analyses for the weak lensing approximation.
In general, the same behavior for the DR approximation was obtained, with weaker constraints to $|\langle\delta\rangle_{1D}|$. This is expected, since this
functional form for $\alpha$ dilutes the effect, which turns out to be more difficult to detect. 

The results for the Union2.1 compilation data \cite{union2.1,rubin} are completely consistent with lines of sight with average 
density, i.e. $|\langle\delta\rangle_{1D}|\sim 0$, although a big region in the parameter space is allowed regardless the value for $H_0$ adopted in the analysis.

As happened for the DR approximation, when considering the SDSS compilation data \cite{sdss}, there is a tension between the samples which is
alleviated with smaller values for $H_0$. The results for the weak lensing approximation are summarized in Table \ref{table2}.

\subsection{The flux-averaging approximation} 
\label{FLUX}

The method described in Sec. \ref{faa} is applied to a flat $\Lambda$CDM model to derive bounds to the matter density parameter $\Omega_{\rm m}$.
We consider a flat $\Lambda$CDM model and marginalize over $H_0$, where the results derived are presented in Fig. \ref{fig3}.

Figure \ref{fig3}(a)-(b) displays the results by considering the Union2.1 compilation data \cite{union2.1,rubin}, $H(z)$ measurements \cite{hzmeasurements} 
and two values for $H_0$ \cite{H0CR,H0riess}. As the number of bins is somehow arbitrary, we choose $\Delta z=0.08$ to guarantee
several supernovae in each bin, which provided 19 bins in total. With the SNe Ia dataset, the matter density parameter is constrained to be
in the interval: $\Omega_{\rm m}=0.26^{+0.08}_{-0.07}$ $(2\sigma)$. One may see that the $H(z)$ measurements with $H_0$(CR) prefer higher values for 
$\Omega_{\rm m}$. When the joint analysis is performed, we get a little improvement in the constraints, where $\Omega_{\rm m}=0.29 \pm 0.06$
within the $2\sigma$ confidence level. When $H_0$(R) is considered, both samples provide basically the same constraints, 
with $\Omega_{\rm m}=0.26^{+0.06}_{-0.05}$ $(2\sigma)$.
Varying the number of supernovae in a bin did not change the results noticeably, with a higher
impact in quality of the fit (minimum $\chi^2$ divided by the number of degrees of freedom).

Figure \ref{fig3}(c)-(d) shows the analyses with the SDSS compilation data \cite{sdss}, $H(z)$ measurements \cite{hzmeasurements} 
and two values for $H_0$ \cite{H0CR,H0riess}. In this case, we set $\Delta z=0.1$ and we have 15 supernovae bins. 
For the SNe Ia sample, the matter density parameter is in the interval: $\Omega_{\rm m}=0.27^{+0.09}_{-0.08}$.
The joint analysis with $H_0$(CR) provides $\Omega_{\rm m}=0.31 \pm 0.06$ $(2\sigma)$, and with $H_0$(R) we have
$\Omega_{\rm m}=0.28^{+0.06}_{-0.05}$ $(2\sigma)$. Again, different values for $\Delta z$ provided essentialy the same results.

Note that in the flux averaging approximation the SDSS sample \cite{sdss} is completely compatible with the $H(z)$ measurements 
\cite{hzmeasurements} regardless the value of $H_0$. Therefore, in this approximation there is no tension as derived for the DR and weak
lensing approximations. Table \ref{table3} synthetizes the results for the flux averaging approximation.

\begin{table}[htbp]
\caption{Limits to $\Omega_{\rm m}$ in the flux averaging approximation.}
\label{table3}
\begin{center}
\begin{tabular}{@{}cccc@{}}
\hline Sample & $\Omega_{\rm m}$ ($2\sigma$) & $N_{bins}$ & $\chi^2_{min}$
\\ \hline\hline
Union2.1 & $0.26^{+0.08}_{-0.07}$  & 19 & $18.0$ \\
Union2.1 $+$ $H(z)$ $+$ $H_0$(CR)   & $0.29^{+0.06}_{-0.06}$ & 19  &  $32.9$ \\
Union2.1 $+$ $H(z)$ $+$ $H_0$(R)  & $0.26^{+0.06}_{-0.05}$ & 19 &  $33.0$ \\
SDSS & $0.27^{+0.09}_{-0.08}$ & 15 & $25.1$ \\
SDSS $+$ $H(z)$ $+$ $H_0$(CR)   & $0.31^{+0.06}_{-0.06}$ & 15  & $53.0$ \\
SDSS $+$ $H(z)$ $+$ $H_0$(R)  & $0.28^{+0.06}_{-0.05}$ & 15 &  $54.1$ \\
\hline
\end{tabular}
\end{center}
\end{table}

\begin{figure*}[ht!]
     \begin{center}
        \subfigure[\, Union2.1 compilation data \cite{union2.1,rubin}.]{%
            \label{fig:first_fig15}
            \includegraphics[width=0.4\textwidth]{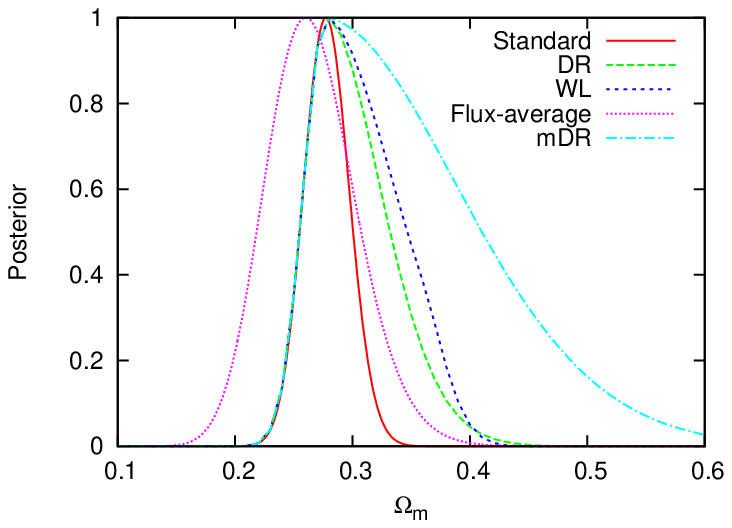}
        }%
        \subfigure[\, SDSS compilation data \cite{sdss}.]{%
           \label{fig:second_fig15}
           \includegraphics[width=0.4\textwidth]{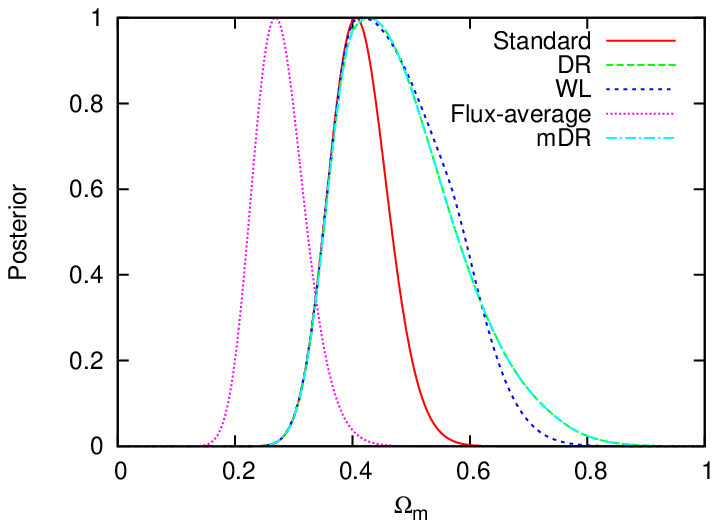}
        }

    \end{center}
    \caption{%
        (color online). \emph{Comparison among different approximations.} Posterior probability for $\Omega_{\rm m}$ from SNe Ia data for different approximations. 
        The solid red line corresponds to a perfect homogeneous universe, the standard case. The dashed green line refers to the DR approximation,
        while the short-dashed blue line corresponds to the weak lensing (WL) approximation. The dotted magenta line shows the posterior for 
        the flux-averaging approximation and the dot-dashed light blue line stands for the modified DR (mDR) approximation.
        (a) The results for the Union2.1 compilation data \cite{union2.1,rubin}. (b) The results for the SDSS compilation data \cite{sdss}.
     }%
   \label{fig5}
\end{figure*}

\subsection{The modified DR approximation} 
\label{MDR} 

Now, we discuss the results from the statistical analyses performed within the modified DR approximation discussed in 
Sec. \ref{mdr}. We marginalize over $H_0$ and we are left with two free parameters, $\Omega_{\rm m}$ and $\alpha_{\rm m}$.

Figure \ref{fig4} displays the results with the Union2.1 compilation data \cite{union2.1,rubin}, 
the SDSS compilation data \cite{sdss}, $H(z)$ measurements \cite{hzmeasurements}, and two priors for $H_0$ \cite{H0CR,H0riess}.
There is a great similarity between the results from the original and the modified DR approaches. As in the modified DR the different expansion
rate can mimick, to a certain extent, the cosmological constant, higher values for $\Omega_{\rm m}$ are allowed. But, at the end of the day,
as the $H(z)$ measurements give an upper constraint to the matter density parameter, the constraints are quite similar in the joint analyses.
In this way, this approximation cannot avoid dark energy, since $\Omega_{\rm m}=1.0$ is excluded with high confidence. As a constant 
$\alpha_{\rm m}$ represents the highest deviation from the standard homogeneous case for this parametrization, no function for $\alpha_{\rm}(z)$ can do the job, because
as the smoothness parameter tends to one for high redshifts, one would get an even smaller deviation from the standard case.

For the smoothness parameter $\alpha_m$ an improvement occurs to the constraints. This is due to the fact that in this approximation $\alpha_m$ plays a more
important role, acting not only in the focusing of the light beam but also changing the local expansion rate. The results are still in full
agreement with a perfect homogeneous universe $(\alpha_m=1.0)$, and within this model the smoothness parameter is more restricted by the
observational data. For the SDSS sample the tension remains, since this approximation forces still higher values for $\Omega_{\rm m}$ compared
to the DR approximation. Therefore, a different local expansion rate along the line of sight is not able to provide a good agreement between the
SDSS compilation data \cite{sdss} and the $H(z)$ measurements \cite{hzmeasurements}. The summary of the constraints for the modified DR are shown in Table \ref{table4}.

\begin{table}[htbp]
\caption{Limits to $\alpha_{\rm m}$ and $\Omega_{\rm m}$ in the modified DR approximation.}
\label{table4}
\begin{center}
\begin{tabular}{@{}cccc@{}}
\hline Sample & $\Omega_{\rm m}$ ($2\sigma$) & $\alpha_{\rm m}$ ($2\sigma$) &
$\chi^2_{min}$
\\ \hline\hline
Union2.1 & $0.28^{+0.23}_{-0.04}$ & $0.99^{+0.01}_{-0.70}$ & $562.2$ \\
Union2.1 $+$ $H(z)$ $+$ $H_0$(CR)  & $0.32^{+0.08}_{-0.06}$  & $0.80^{+0.20}_{-0.33}$ & $576.1$ \\
Union2.1 $+$ $H(z)$ $+$ $H_0$(R) & $0.28^{+0.07}_{-0.03}$ & $0.99^{+0.01}_{-0.35}$ & $577.2$ \\
SDSS & $0.42^{+0.58}_{-0.10}$ & $0.95^{+0.05}_{-0.84} $  & $154.8$ \\
SDSS $+$ $H(z)$ $+$ $H_0$(CR)  &$0.37^{+0.07}_{-0.06}$  & $1.0^{+0.0}_{-0.28}$     &$169.9$ \\
SDSS $+$ $H(z)$ $+$ $H_0$(R) & $0.33^{+0.06}_{-0.05}$ & $1.0^{+0.0}_{-0.20}$ & $174.4$ \\
\hline
\end{tabular}
\end{center}
\end{table}

\subsection{Discussion}

Now we compare how the different approximations impact other cosmological parameters, in our case $\Omega_{\rm m}$. 

Figure \ref{fig5} shows the results for the four different approximations and also when the inhomogeneities are neglected, the standard case.
In Fig. \ref{fig:first_fig15} the results for the Union2.1 compilation data \cite{union2.1,rubin} are shown. The behavior for the DR, 
the weak lensing, and the modified DR approximations are expected. We see that all of them allow higher values for $\Omega_{\rm m}$. As 
the combination $\alpha \Omega_{\rm m}$ is constrained, considering values such that $\alpha <1$ implies higher values for $\Omega_{\rm m}$.
On the other hand, the flux-averaging approximation prefers lower values for $\Omega_{\rm m}$. In general, regardless the approximation considered,
the posteriors are broader, which implies larger errors in $\Omega_{\rm m}$. These errors are not negligible, but do not put a severe problem to 
the standard model at this moment. However, they can have a larger impact in joint analyses, since the intersection in the parameter space
may be different.

Figure \ref{fig:second_fig15} shows the comparison for the different approximations with the SDSS compilation data \cite{sdss}. The DR, the weak
lensing, and the modified DR approximations showed the same pattern, where higher values for $\Omega_{\rm m}$ are compatible with the data compared to the standard
case. In the case of the flux-averaging approximation, much lower values were derived compared to the Union2.1 compilation data \cite{union2.1,rubin}.
It is this shift which turns possible that this approximation to be compatible with the $H(z)$ measurements \cite{hzmeasurements}.

While the effects discussed here cannot change our whole view about the way cosmological data are interpreted, the trend is they become 
more and more important due to several planned and ongoing surveys which will discover thousands of supernovae.

\section{Conclusions}
\label{sec_conc}

We have investigated the effects of small-scale inhomogeneities in light propagation of narrow beams in different approximations. We have studied four
approximations, the DR approximation \cite{dr}, the weak lensing approximation \cite{bol_dr_2011}, the flux-averaging approximation \cite{flux_av},
and a modified DR approximation. We used the 581 SNe Ia from the Union2.1 compilation data \cite{union2.1,rubin}, 288 SNe Ia from
the SDSS compilation data \cite{sdss}, 19 $H(z)$ measurements and two values for $H_0$ \cite{H0CR,H0riess}.

For the Union2.1 compilation data \cite{union2.1,rubin} all approximations are in agreement with observational data. Basically, the approximations incur in higher errors for the
matter density parameter $\Omega_{\rm m}$. Although the difference from the perfectly homogeneous case is not catastrophic, the effects
must be important in order to achieve accurate results.

On the other hand, the 288 SNe Ia from the SDSS compilation data \cite{sdss} prefers more exotic models compared to $\Lambda$CDM \cite{exotic}.
So we have analyzed the effects of the inhomogeneities in this sample together with $H(z)$ measurements \cite{hzmeasurements} and two values
for $H_0$ \cite{H0CR,H0riess} in order to see whether a better agreement between different samples can be obtained. For the DR, the weak
lensing, and the modified DR approximations the tension remains, although it can be slightly alleviated for higher values for $H_0$. However, as the flux-averaging
approximation provides lower values for $\Omega_{\rm m}$, the tension disappears irrespective of the value of $H_0$ considered in the 
marginalization.

In summary, the non-linear effects of the inhomogeneities in narrow beams, in addition to bias and selection effects may play an important role in the estimation
of cosmological parameters, as well as the determination of the nature of dark energy or a need to introduce a different gravitational theory on 
large scales. Therefore, efforts in modelling such effects together with observational scrutiny are vital to build a consistent picture
of the Universe.

\vspace{1.0cm}

\noindent{\bf Acknowledgements:}  V.C.B is supported by CNPq-Brazil through a fellowship within the programme Science without Borders. R.F.L.H thanks INCT-A and is supported by CNPq (No. 478524/2013-7).


\begin{thebibliography}{99}
\bibitem{other} The approximations studied here are not exhaustive, for other approaches,  see e.g. \cite{alpham1,other_app,other_app2}.
\end{thebibliography}
\end{document}